\documentclass{aastex631}
\usepackage{amsmath}
\DeclareUnicodeCharacter{2212}{-}

\begin{document}

\title{Unraveling the Thermodynamic Enigma between Fast and Slow Coronal Mass Ejections}

\correspondingauthor{Soumyaranjan Khuntia, Wageesh Mishra and Yuming Wang}
\email{soumyaranjan.khuntia@iiap.res.in, wageesh.mishra@iiap.res.in, ymwang@ustc.edu.cn}

\author[0009-0006-3209-658X]{Soumyaranjan Khuntia}
\affiliation{Indian Institute of Astrophysics, II Block, Koramangala, Bengaluru 560034, India}
\affiliation{Pondicherry University, R.V. Nagar, Kalapet 605014, Puducherry, India}

\author[0000-0003-2740-2280]{Wageesh Mishra}
\affiliation{Indian Institute of Astrophysics, II Block, Koramangala, Bengaluru 560034, India}

\author[0000-0003-2129-5728]{Sudheer K Mishra}
\affiliation{Indian Institute of Astrophysics, II Block, Koramangala, Bengaluru 560034, India}
\affiliation{Astronomical Observatory, Kyoto University, Sakyo, Kyoto 606-8502, Japan}

\author[0000-0002-8887-3919]{Yuming Wang}
\affiliation{CAS Key Laboratory of Geospace Environment, Department of Geophysics and Planetary Sciences, University of Science and Technology of China, Hefei 230026, People's Republic of China}

\author[0000-0003-0951-2486]{Jie Zhang}
\affiliation{Department of Physics and Astronomy, George Mason University, 4400 University Dr., MSN 3F3, Fairfax, VA 22030, USA}

\author[0000-0002-2349-7940]{Shaoyu Lyu}
\affiliation{CAS Key Laboratory of Geospace Environment, Department of Geophysics and Planetary Sciences, University of Science and Technology of China, Hefei 230026, People's Republic of China}


\begin{abstract}
Coronal Mass Ejections (CMEs) are the most energetic expulsions of magnetized plasma from the Sun that play a crucial role in space weather dynamics. This study investigates the diverse kinematics and thermodynamic evolution of two CMEs (CME1: 2011 September 24 \& CME2: 2018 August 20) at coronal heights where thermodynamic measurements are limited. The peak 3D propagation speed of CME1 is high (1,885 km s$^{-1}$) with two-phase expansion (rapid and nearly constant), while the peak 3D propagation speed of  CME2 is slow (420 km s$^{-1}$) with only a gradual expansion. We estimate the distance-dependent variations in the polytropic index, heating rate, temperature, and internal forces implementing the revised FRIS model, taking inputs of 3D kinematics estimated from the GCS model. We find CME1 exhibiting heat-release during its early-rapid acceleration decrease and jumps to the heat-absorption state during its constant acceleration phase. In contrast to CME1, CME2 shows a gradual transition from the near-adiabatic to the heat-absorption state during its gradually increasing acceleration. Our analysis reveals that although both CMEs show differential heating, they experience heat-absorption during their later propagation phases, approaching the isothermal state. The faster CME1 achieves an adiabatic state followed by an isothermal state at smaller distances from the Sun than the slower CME2. We also find that the expansion of CMEs is primarily influenced by centrifugal and thermal pressure forces, with the Lorentz force impeding expansion. Multi-wavelength observations of flux-ropes at source regions support the FRIS model-derived findings at initially observed lower coronal heights. 
\end{abstract}



\section{Introduction} \label{sec:intro}

Coronal mass ejections (CMEs) are massive magnetic structures originating from the Sun and extending into the heliosphere \citep{Hundhausen1984,Chen2011,Webb2012}. Due to their ability to cause intense geomagnetic storms and impact space weather, CMEs present a significant concern for society's reliance on space-based infrastructure \citep{Gonzalez1994,Pulkkinen2007}. Moreover, understanding CME initiation, propagation, and internal thermodynamics is crucial for advancing knowledge about the plasma properties of magnetised plasmoids from a scientific standpoint. Previous studies on CMEs have primarily focused on their source region dynamics, kinematics, arrival times, in-situ signatures, and geo-effectiveness \citep{Harrison1995, Webb2000, Gopalswamy2009,Mishra2013,Lugaz2017}. Extensive research has also been conducted on the kinematic evolution of interacting CMEs \citep{Gopalswamy2001,Shen2012,Mishra2014, Mishra2015, Mishra2017}. However, our understanding of the internal thermodynamic properties of CMEs, during their evolution from near to farther distances from the Sun is very limited. 

The underlying mechanism for global acceleration and hence the kinematics of CMEs depend on their internal thermodynamic properties. A better understanding of the internal properties will enable the estimation of kinematics with greater accuracy, which will help not only in forecasting the arrival time but also give a better assessment of the geo-effectiveness. There are studies using spectroscopic observations from the UltraViolet Coronagraph Spectrometers (UVCS), Coronal Diagnostic Spectrometer (CDS), and Solar Ultraviolet Measurements of Emitted Radiation (SUMER) instruments on the \textit{SOlar and Heliospheric Observatory} (SOHO) \citep{Domingo1995} which have provided insights into the density, Doppler velocity, temperature, and ionization state of CMEs \citep{Raymond2002,Kohl2006,Lee2009,Bemporad2022}. The heating of CMEs has been reported using spectroscopic measurements of the erupting prominence material \citep{Filippov2002,Lee2017} and investigating the ionisation states at 1AU \citep{Rakowski2007}. The spectroscopic diagnosis of plasma temperature, density, and heating of CMEs at few solar radii from the Sun suggests that CMEs are loop-like structures with temperatures higher than the typical corona near the Sun \citep{Ciaravella2000,Ciaravella2003, Kohl2006}. Using measurements from UVCS, it was found that as the CME travels $3.5 R_\odot$, the total thermal energy gained is comparable to the kinetic and gravitational potential energies \citep{Akmal2001}. Further, the in-situ signature of ICMEs shows lower temperatures, stronger magnetic fields \citep{Burlaga1981,Liu2006,Richardson2010,Kilpua2017}, and higher charge ion states \citep{Lepri2001,Zurbuchen2006} compared to the surrounding solar wind medium. Using in-situ observations of CMEs at various distances from the sun, it is established that density and magnetic field decrease faster in ICMEs than in Solar wind; however, temperature decreases slower in ICMEs than in solar wind \citep{Wang2004a,Liu2006,Li2016}. These findings are based on estimating the polytropic index of CME plasma, which considers a polytropic process that can describe the evolution of a CME from one thermodynamic state to another. This suggests a specific relation between plasma pressure and density ($ p\propto \rho^\Gamma $) exists during the evolution of CMEs. Using multispacecraft in-situ observations between 0.3-20 AU, the polytropic index of CMEs was found to be around 1.1-1.3, which suggests the expansion of the CMEs is closer to the isothermal state \citep{Osherovich1993,Liu2005,Liu2006}.

Earlier studies have characterised different space plasma by different polytropic indices ranging from sub-adiabatic to super-adiabatic values. For example, super-adiabatic values of polytropic indices are reported in coronal plasma, Earth's plasma sheet, planetary magnetosphere, and even galaxy clusters \citep{Borovsky1998,Tatrallyay1984,Bautz2009}. In contrast, solar wind plasma observed at sub-AU and AU distances from the Sun shows a positive correlation between pressure and density, with an average polytropic index ranging from 1.4-1.6, close to adiabatic process \citep{Totten1995,Newbury1997,Livadiotis2016}. The plasma in the outer heliosphere, inner heliosheath, magnetosheath, and bow shock is characterised by a negative correlation between their pressure and density, implying a polytropic index lesser than unity \citep{Livadiotis2011,Nicolaou2015}. These studies indicate the importance of estimating the polytropic index for characterizing the space plasma's thermodynamic state and heat flux in general without dealing with complex energy equations. Further recent studies have examined the polytropic behaviour of different substructures of ICMEs, such as turbulent sheath and quite magnetic ejecta region \citep{Dayeh2022}. In this study, both the sheath and magnetic ejecta are much away from isothermal behaviour, but the ejecta is much closer to adiabatic. Attempts have also been made to develop and implement a theoretical model on the solar wind in-situ observations at 1 AU to understand the effect of temperature anisotropy on the adiabatic and non-adiabatic polytropic index \citep{Livadiotis2021}. These studies, primarily utilizing in-situ observations of ICMEs, provided some insights into the internal thermodynamic properties but were only limited to certain heliocentric distances or at a particular time during the heliospheric propagation of the ICMEs. Therefore, it is evident that the continuous estimation of the polytropic index of CME during their outward journey from near the Sun to beyond is less explored.

The lack of understanding of the evolution of internal thermodynamics of CMEs is partly due to observational limitations. The spectroscopic observations can provide thermal information of CMEs, but such measurements are limited to only near the Sun \citep{Antonucci1997,Bemporad2010}. The temperature of ICMEs and inference of its thermal state is possible using in-situ observations, which only provide measurements at certain heliocentric distances from the Sun \citep{Zurbuchen2006,Richardson2010}. Further, due to the sparse distribution of in-situ spacecraft, it is difficult for multiple spacecraft to get co-aligned to measure the thermal state of the same ICME at different distances \citep{Phillips1995,Winslow2021}. It is often challenging to associate the remote sensing observations of global structures of CMEs with their local in-situ observations \citep{Mishra2013,Mishra2023}. Most of the routine coronagraphic observations of CMEs are in white-light, which does allow continuous tracking of density-enhanced structures, but lacks any thermodynamic information. As one of the three-part structures seen in white-light observations of CMEs, the dark cavity is often associated with magnetic flux-rope (MFR) observed in multi-wavelength observations \citep{Gibson2006,Riley2008}. The MFR-associated CMEs are often identified as magnetic clouds in the in situ observations \citep{Richardson2010,Wang2016}. Using multi-wavelength imaging observations and the differential emission measure (DEM) technique \citep{Cheung2015}, it is possible to infer the thermal behaviour of MFR at the lower coronal height \citep{Cheng2012,Gou2019,Sheoran2023}. The white-light observations of CMEs, capable of providing their kinematics, if combined with appropriate models, can probe the thermodynamics of CMEs; both the dynamic and thermal properties are essentially interlinked. Therefore, using multi-wavelength observations of MFRs, their white-light observations, and in-situ observations together should be extensively studied.

The efforts to understand the thermodynamic evolution of CMEs from a modelling perspective is not much undertaken except a handful of case studies using the flux-rope internal state (FRIS) model \citep{Wang2009,Mishra2018,Mishra2020}. Also, most of the MHD models simulating the kinematic behaviour of CMEs adopt an ad-hoc value of the polytropic index. Most of these models use the polytropic index value lesser than 5/3 in their global MHD simulations of ICMEs \citep{Riley2003,Manchester2008,Mayank2022}. Thus, there is a lack of estimates of the internal thermodynamics parameters of the CMEs in the interplanetary space between the Sun and 1AU. It is imperative to probe the thermal state of CMEs and understand the physical mechanism for their heating in interplanetary space.

In the present work, we investigate the continuous evolution profile of the internal thermal properties of CMEs using their global kinematics. For this purpose, we first make necessary modifications to rightly advance the existing analytical flux-rope Internal State (FRIS) model \citep{Wang2009,Mishra2018,Mishra2020}, which has earlier been implemented in only a few case studies. FRIS model can derive the various thermodynamic parameters, e.g., polytropic index, temperature, heating rate, and different internal forces acting on the CME flux-rope. We implemented the revised FRIS model to two extreme cases of CMEs; one very fast-speed CME of 2011 September 24 (hereinafter, CME1) and another slow-speed CME of 2018 August 20 (hereinafter, CME2). The selected CME1 and CME2 have been previously reported in the literature for their unique kinematics and geoeffective response \citep{Wood2016,ChinChun2016,Chen2019,Palmerio2022}. We focus on understanding the different thermal histories of the selected CMEs with significantly different speeds and accelerations profiles for both propagation and expansion. Using multi-wavelength observations, we also investigate the evolution of flux-rope proxies of the selected CMEs at the source regions and in the lower corona. Such studies have the potential to suggest putting further constraints in different MHD models dealing with the propagation of CMEs in the ambient solar wind and their arrival time at Earth.

\section{Methodology}
\label{sec:methods}

\begin{figure}[ht]
\centering
\includegraphics[width=0.55\linewidth]{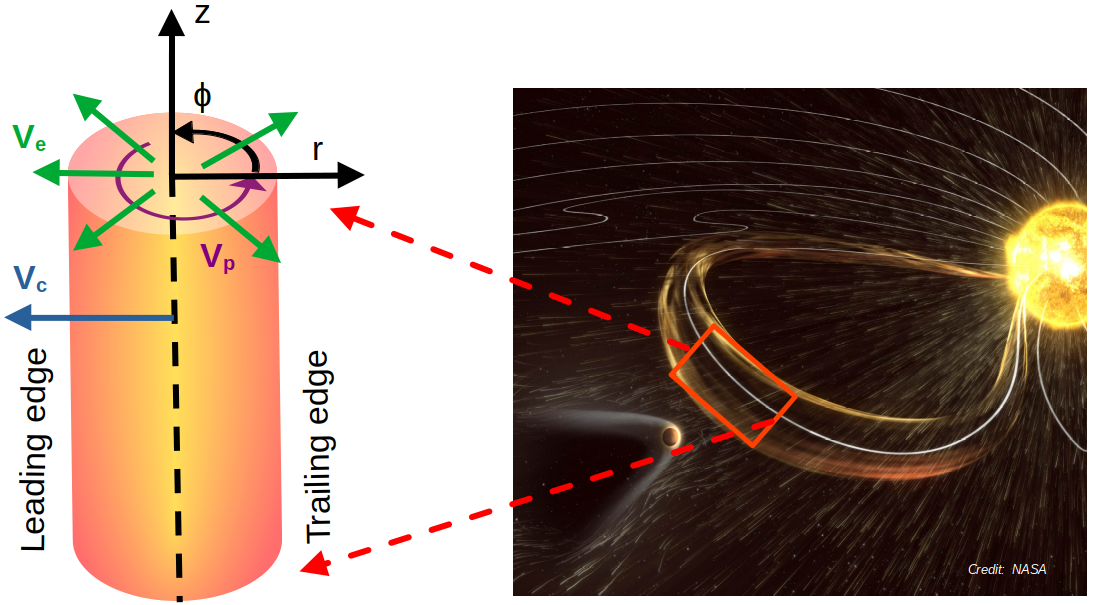}
\caption{Schematic of a ﬂux-rope CME in the cylindrical coordinate system (i:e r,$\phi$, z) showing the propagation speed ($v_c$ ) of the axis of the flux-rope, expansion speed ($v_e$ ), and poloidal speed ($v_p$ ).}.
\label{fig:fris}
\end{figure}

The development of the FRIS model and its subsequent revision is described earlier \citep{Wang2009,Mishra2018}; however, there are miscalculations in the derivation of some equations in \citet{Mishra2018} which recently have been corrected in \citet{Mishra2023b}. Here we briefly reiterate the important points of the revised model for completeness. The model explores the heliospheric propagation of a coronal mass ejection (CME) as it expands in a cylindrical shape at a local scale (Figure \ref{fig:fris}) with mass and angular momentum being conserved. It assumes a self-similar expansion and considers three global motions: linear propagation speed ($v_c$) of the axis, expansion speed ($v_e$), and poloidal speed ($v_p$) of the CME flux-rope. The model focuses on the radial expansion of the flux-rope, considering the magnetohydrodynamic equation, thermal pressure force, Lorentz force from the axis to the boundary of the flux-rope, and centrifugal force resulting from the poloidal motion of the plasma. The FRIS model offers a new approach to studying CME's internal state by examining the relationships between macroscopic kinetic and thermodynamic parameters, considering a polytropic process for CME evolution. The detailed derivation of the FRIS model can be found in the appendix (\ref{appendix:derivation}).

The final equation of motion for the radial expansion of the CME flux-rope can be written in terms of the measurable kinematic parameters such as the distance (L) of the center of the CME flux-rope from the surface of the Sun, the radius of the flux-rope (R), and their time derivatives as,

\begin{align}
\frac {R}{L} =  &{c_5 \biggl[\frac {a_e R^2}{L}\biggr]}-{c_3 c_5 \biggl[\frac {R}{L^2}\biggr]}-{c_2 c_5\biggl[\frac {1}{R}\biggr]}-{c_1 c_5\biggl[\frac {1}{LR}\biggr]}{\nonumber}\\
&+c_4 \biggl[\frac {da_e}{dt}+{\frac {(\gamma -1)a_e v_c}{L}}+\frac{(2\gamma -1)a_e v_e}{R}\biggr] + c_3 c_4\biggl[ \frac{(2-\gamma) v_c }{L^2 R}+{\frac{(2-2\gamma) v_e }{LR^2}}\biggr] {\nonumber}\\ 
&+c_2 c_4\biggl[{\frac{(4-2\gamma) v_e L}{R^4}}-{\frac{\gamma v_c}{R^3}}\biggr] + c_1 c_4\biggl[{\frac{(4-2\gamma) v_e}{R^4}}+{\frac{(1-\gamma)v_c}{LR^3}}\biggr]
\label{eqn:fitting1}
\end{align}
\noindent
Where $\gamma$ is the adiabatic index and $c_1 - c_5$ are unknown constants, whose values can be obtained by fitting R.H.S to L.H.S of the Equation (\ref{eqn:fitting1}). The derived internal dynamic and thermodynamic parameters are summarized in Table (\ref{tab:parameters}).
\begin{table*}[ht]
\caption{\label{tab:parameters} List of the Derived Internal Thermodynamic Parameters from the FRIS Model. The details of coefficients ($c_{1-5}$) and factors ($k_{1-11}$) are the same as in Table 1 of \citet{Mishra2018}.}
\hspace{15pt}
\begin{tabular}{lccc}
\hline
\textbf{Quantities} & \textbf{Factors} & \textbf{Values} & \textbf{SI Units} \\[3pt]
\hline
\hline
Lorentz Force (${{\bar f}_{em}}$) & \(\frac{ k_2 M}{k_7 }\) & \( { {c_2 R^{-5}}+{c_3 L^{-2} R^{-3}} }\) &$Pa.m^{-1}$  \\[3pt]
Thermal pressure Force (${{\bar f}_{th}}$) & \( \frac{ k_2 M}{k_7 }\) & \( { \lambda(t) L^{-\gamma} R^{-2\gamma -1} }\) &$Pa.m^{-1}$  \\[3pt]
Centrifugal Force (${{\bar f}_{p}}$) & \( \frac{ k_2 M}{k_7 }\) &\( { c_1 R^{-5} L^{-1} }\)& $Pa.m^{-1}$  \\[3pt]
Proton number density (${{\bar n}_{p}}$) & \( \frac{ M}{k_7 }\) & \( \frac{1}  {\pi  m_p} ( L R^2 )^{-1}\)& $m^{-3}$ \\[3pt]
Thermal pressure ($\bar {p}$) & \( \frac{ k_2 k_8 M}{k_4 k_7 }\) &\( { \lambda (LR^2)^{-\gamma} }\) & $Pa $\\[3pt]
Temperature ($\bar {T}$) & \( \frac{ k_2 k_8 }{k_4 }\) & \( \frac{\pi \sigma}   {(\gamma -1)} {\lambda (LR^2)^{1{-\gamma}}  }\) &$ K $\\[3pt]
Rate of change of entropy ($\frac{ds}{dt}$) & & \( \frac{1}{{\sigma \lambda}}\frac{d\lambda }{dt }\) & $J.K^{-1}.kg^{-1}.s^{-1}$ \\[3pt]
Heating rate ($\bar{\kappa}$) & \( \frac{ k_2 k_8 }{k_4 }\) & \( \frac{\pi }   {(\gamma -1)} { (LR^2)^{1{-\gamma}}  }\frac{d\lambda }{dt }\) & $J.kg^{-1}.s^{-1}$ \\[3pt]
Polytropic Index ($\Gamma$) & & \(   \gamma +  \displaystyle{ \frac{ln{\frac{\lambda (t)}  {\lambda (t+\Delta t)}}}  {ln[(\frac{L(t+\Delta t)}  {L(t)})[\frac{R(t+\Delta t)}  {R(t)}]^2]} }\) &  \\[3pt]
\hline
\end{tabular}

\end{table*}

As the FRIS model takes the kinematics as input to derive the internal parameters, we have used the continuous coronagraphic data to track the selected CMEs away from the Sun. The coronagraphic observations of CME suffer from projection effects as they capture the two-dimensional POS images of three-dimensional (3D) structures. For both the selected CME1 and CME2, being Earth-directed, they will experience maximum projection effects from SOHO/LASCO observations, i.e., coronagraphs located at L1 will underestimate the speed while angular width will be overestimated. To estimate the 3D kinematics of CMEs, one needs to exploit coronagraphic observations from multiple viewpoints and 3D reconstruction methods \citep{Mierla2010,Davies2013,Mishra2014}. For this purpose, we have used the Graduated Cylindrical Cell (GCS) model \citep{Thernisien2006,Thernisien2011} and estimated 3D kinematics, where we can find out the positional and geometrical parameters of CME. The GCS forward modelling method has been regularly used to determine the 3D kinematic parameters of the flux-rope CMEs \citep{Vourlidas2013,Wang2014,Mishra2015}.

\section{Results}\label{sec:results}

\subsection{Measurements of CMEs Kinematics from Coronagraphic Observations}
\label{sec:results_CORONA}

To implement the FRIS model on the observations of selected CMEs, we use the white-light coronagraphic observations to estimate the kinematics of the CMEs. The first selected CME (CME1: 2011 September 24) was observed by multiple coronagraphs onboard SOHO and twin \textit{Solar TErrestrial RElations Observatory} (STEREO) \citep{Kaiser2008} spacecraft. The SOHO/LASCO-C2 coronagraph first observed the CME1 at 12:48 UT as a full halo Earth-directed CME with a plane-of-sky (POS) linear speed of 1,915 km s$^{-1}$. The ICME associated with this CME was found to arrive at Earth on 2011 September 26, at 11:34 UT and caused a geomagnetic storm with a Dst minimum reaching -118 nT \citep{Wood2016}. This CME/ICME has been studied for its source regions, radio bursts, and geoeffective properties \citep{Wood2016,ChinChun2016,Hongyu2018}. Another selected CME (CME2: 2018 August 20) of our study was observed by coronagraphs onboard SOHO and STEREO-A spacecraft, as STEREO-B has been unavailable since 2014. The SOHO/LASCO-C2 observed the CME2 first at 21:24 UT, and its projected POS linear speed was 126 km s$^{-1}$. The CME2 arrived at 1 AU on 2018 August 25, 02:00 UT, and was responsible for the third largest geomagnetic storm of solar cycle 24 with a Dst minimum of −174 nT \citep{Chen2019,Gopalswamy2022}. This CME/ICME is also studied extensively for its source regions, interplanetary kinematics, and unusual geo-effectiveness \citep{Mishra2019,Gopalswamy2022,Palmerio2022}. 

\begin{figure*}[ht]
        \centering
        \includegraphics[width=0.42\textwidth,angle=-90]{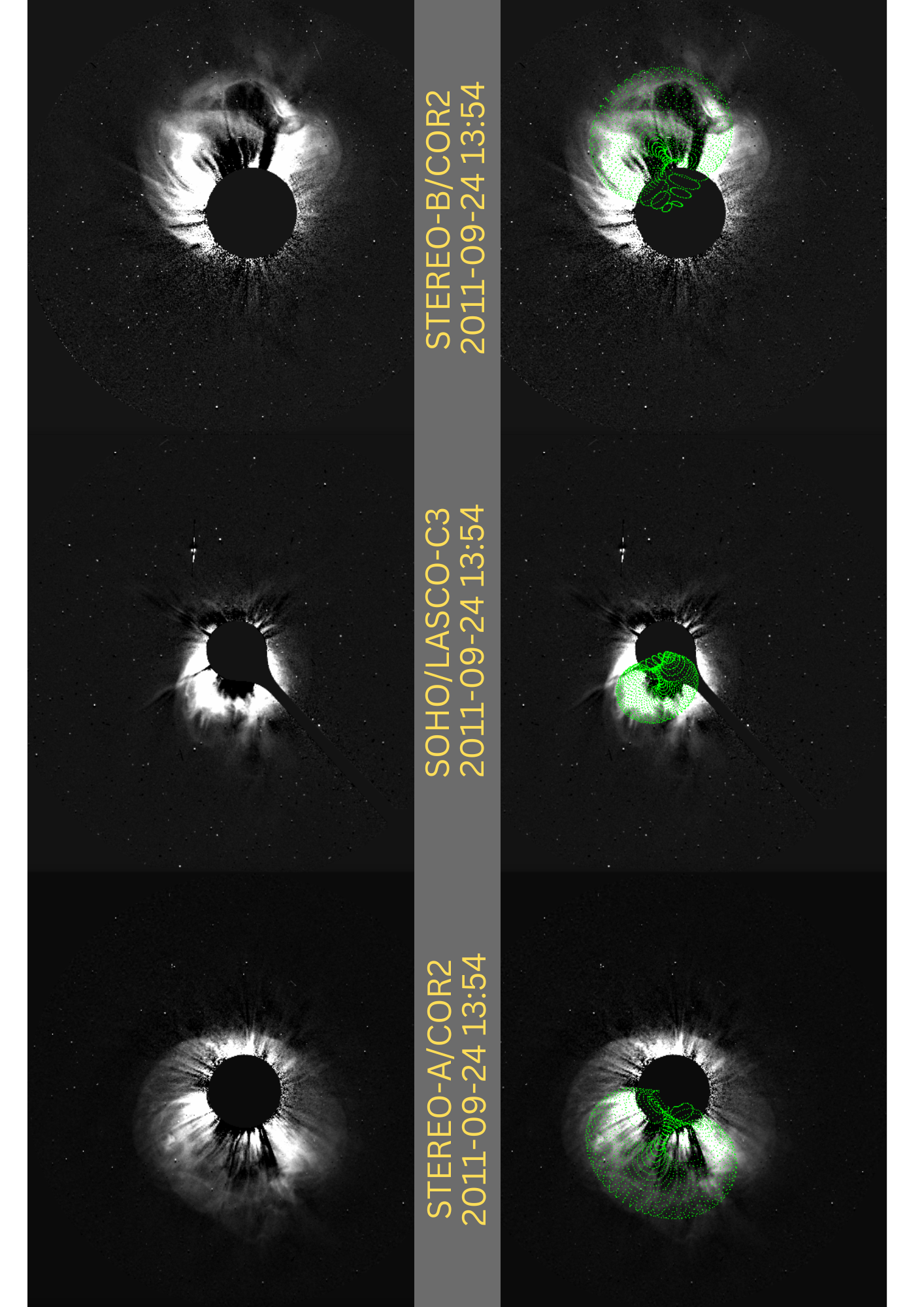}    
        \includegraphics[width=0.42\textwidth,angle=-90]{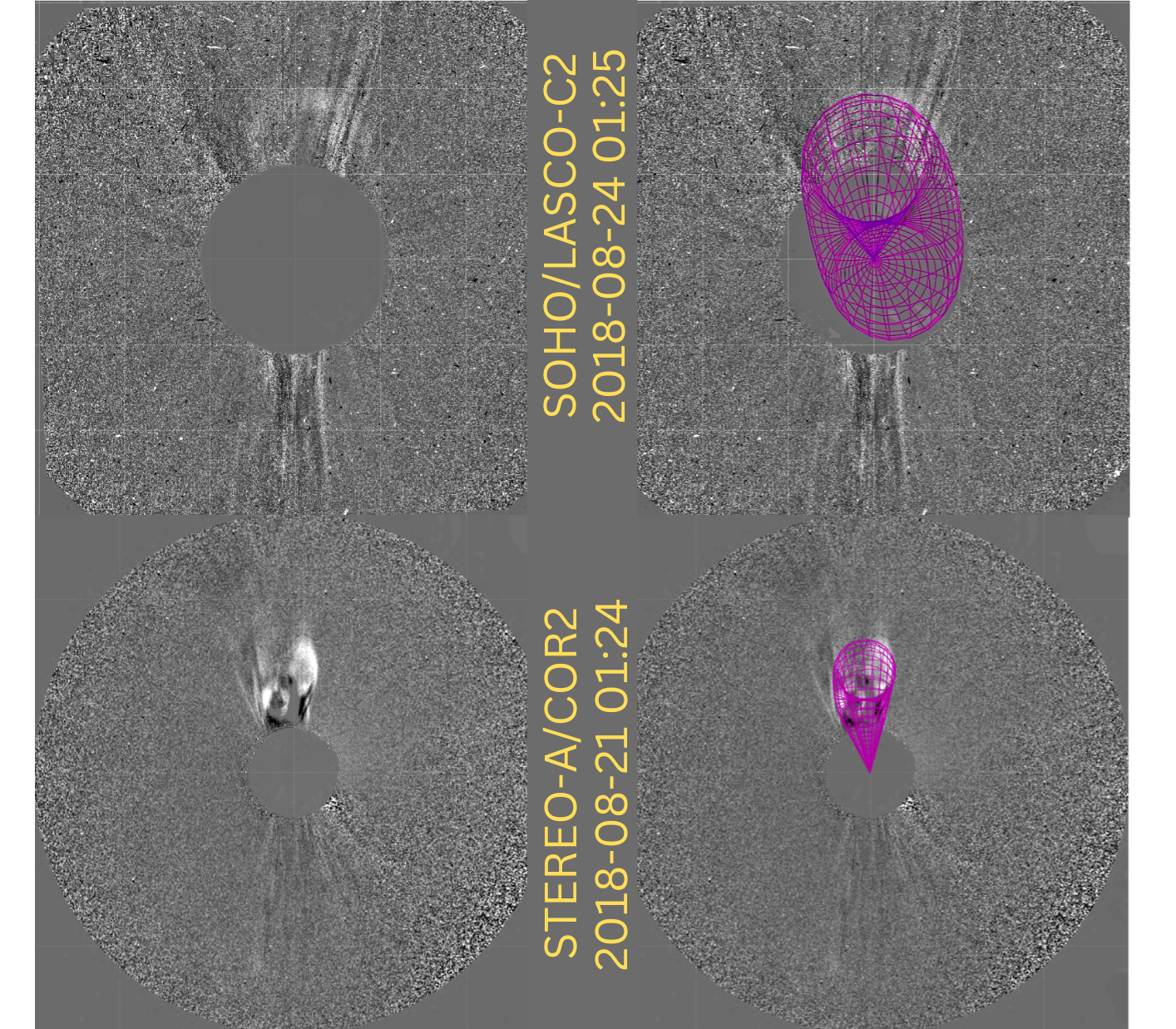}
        \caption{The GCS model fitted wireframe in green and pink overlay on the contemporaneous coronagraphic images of the CME1 shown in left (left: STEREO-A/COR2, centre: LASCO-C3, right: STEREO-B/COR2) and for CME2 shown in the right (left: STEREO-A/COR2, right: LASCO-C2), respectively.}
        \label{fig:coronagraph_images}
\end{figure*}

\subsubsection{3D Kinematic Parameters of CMEs}\label{sec:results_kinema}

We implemented the GCS model to the simultaneous images of CMEs from SOHO/LASCO (C2 \& C3) and STEREO/COR (COR1 \& COR2), as shown in Figure (\ref{fig:coronagraph_images}). However, we could not use STEREO-B/COR observations for CME2 due to the unavailability of data since 2014. Since the GCS model has six free parameters, it is a usual practice to use some of the GCS model parameters from the multi-wavelength observations of the CME source regions \citep{Mostl2014,Palmerio2018}. This helps to reduce the degeneracy in GCS model parameters while we adjust them to mimic the observed CME structure in the coronagraphic observations. In our study, we used multi-wavelength observation of the CME source region to get a rough value of the longitude and latitude to start the model-fitting. It is also possible to derive the flux rope's tilt value by analyzing the orientation of the polarity inversion line and arcade structure at the CME source region \citep{Marubashi2015,Palmerio2018}. It is known that CME suffers maximum deflection and rotation near the Sun; Thus, the initial values of longitude, latitude, and tilt of the CME derived from its source region may vary as CME evolves in the outer corona. Therefore, we did not derive tilt information from the CME source region but relied on the GCS model fitting to the images of CMEs simultaneously taken from the three viewpoints. The GCS model fitted parameters for both CME1 and CME2, at a particular instant, are shown in Table ({\ref{tab:GCS}}). The whole evolution of the model-fitted parameters for each successive time step during our observation is included in the appendix (\ref{appendix:gcs_parameters}).

\begin{table*}[ht]
\caption{\label{tab:GCS} The GCS model fitted geometrical and positional parameters for CME1 (2011 September 24) and CME2 (2018 August 20).}
\begin{tabular} {lccccccc}
\hline
\textbf{Events} & \textbf{Time} & \textbf{Height}  & \textbf{Longitude} & \textbf{Latitude} & \textbf{Aspect}  & \textbf{Tilt}   & \textbf{Half} \\
 & \textbf{(UT)} &   \textbf{($R_\odot$)} & \textbf{($^{\circ}$)} & \textbf{($^{\circ}$)} & \textbf{Ratio} & \textbf{Angle} & \textbf{Angle}\\
  & & & & & &\textbf{($^{\circ}$)} &\textbf{($^{\circ}$)}\\
\hline
\hline
2011 September 24 & 13:54 & 12.9 & -41 & 13 & 0.39 & -62 &26\\
\hline
2018 August 20 & 1:24 & 8.5 & 10 & 5 & 0.27 & 10 & 16\\
\hline

\end{tabular}
\end{table*}

\begin{figure*}[ht]
\centering
        \includegraphics[width=0.43\textwidth,trim={6cm 0cm 22cm 0cm}]{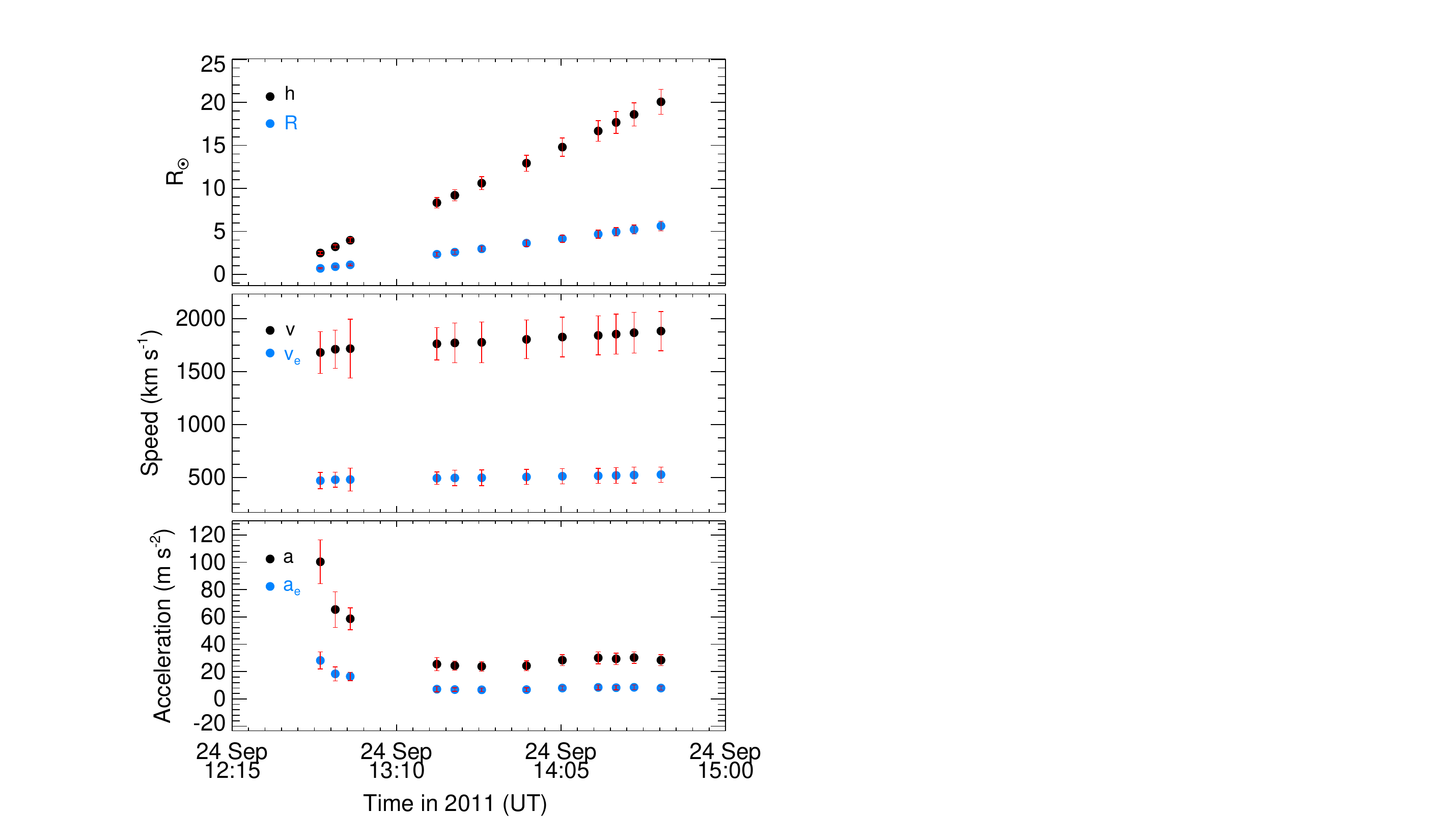}
        \includegraphics[width=0.43\textwidth,trim={4cm 0cm 24cm 0cm}]{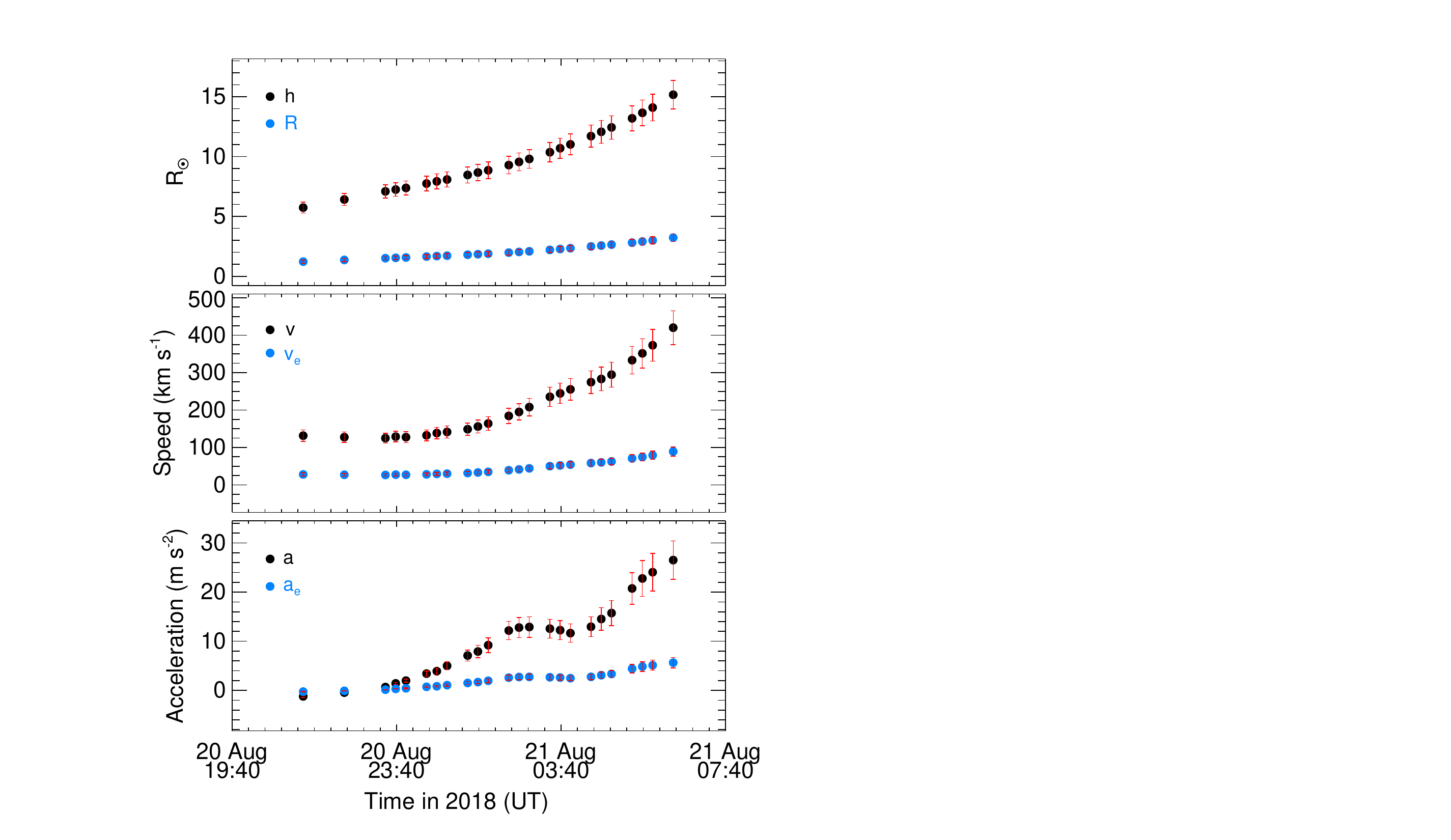}     
        \caption{Kinematics of CME1 (left) and CME2 (right) using the GCS model on the coronagraphic observations. Top panel: the measurements of the heliocentric distance ($h$) of the leading edge of the flux-rope (FR) and its radius ($R$). Middle panel: the propagation speed ($v$) and expansion speed ($v_e$) derived by taking the three-point derivatives of $h$ and $R$, respectively. Bottom panel: the propagation acceleration ($a$) and expansion acceleration ($a_e$) derived by taking the derivative of $v$ and $v_e$, respectively. The red vertical lines at each data point show the error bars derived by considering an error of 10{\%} in the measurements of the flux-rope's leading edge height ($h$).}
        \label{fig:kinematics}
\end{figure*}

The GCS model-derived latitudes of both CMEs suggest that they are propagating in the ecliptic plane. The longitude of CME1 shows its eastward propagation, around 41$^{\circ}$ away from the Sun-Earth line, while CME2 is 10$^{\circ}$ westward from the Sun-Earth line. We also find that the aspect ratio of CME1 is about 44\% larger than that of CME2, while the half angular width of CME1 is only $\approx$ 60\% larger than CME2. Earlier studies have found a positive correlation between CMEs' angular width and their radial POS speeds \citep{Gopalswamy2009}. The GCS model-fitted half angle and aspect ratio remained the same during our observation phase, which satisfied the consideration of the FRIS model that the CME flux rope expands self-similarily. The leading edge height ($h$) of the selected CMEs is estimated from the GCS model fit. Further, the radius of CME flux-rope ($R$) can be estimated as $R = (\frac{\kappa}{1+\kappa})h$, where $\kappa$ is the aspect ratio as derived from the GCS model. We have determined the propagation speed of the CME leading edge and the expansion speed of the CME flux-rope by taking the time derivative of the $h$ and $R$, as shown in Figure (\ref{fig:kinematics}). We used a running three-point box over data points of $h$ and assumed a linear fit to $h$ and $t$ to estimate the derivative at the second point in the box. In contrast, the derivatives at the extreme ends are found out by two-point derivation. This method allows us to visualize the real variation in the speed and acceleration, and will not reduce the data points in the derivatives. 

We have tracked the CME1 and CME2 leading edge up to the height $h\approx20 R_\odot$ and 15 $R_\odot$, respectively. At this maximum height, the radius of the flux-rope $R$ reaches about 5.6 $R_\odot$ and 3.2 $R_\odot$ for the CME1 and CME2, respectively. For the CME1, the leading edge speed $v$ increases up to 1,885 km s$^{-1}$ while the expansion speed $v_e$ reaches about 528 km s$^{-1}$. Similar to our finding, it is reported that CMEs reach their peak speed in low to middle corona \citep{Zhang2004,Temmer2010}. In contrast, CME2 shows a significantly different speed profile where the maximum leading edge speed is about 420 km s$^{-1}$, with a maximum expansion speed of about 90 km s$^{-1}$. Thus it is clear that the CME1 is a fast CME while the CME2 is a slow CME from both propagation and expansion prospective. We note that  CME1 shows a strong decrease in acceleration at the beginning that rapidly decreases in strength up to 8.4 $R_\odot$ followed by a nearly constant acceleration phase as shown in Figure (\ref{fig:kinematics}). Earlier studies have shown that CMEs experience stronger initial acceleration during a shorter time interval within 3 $R_\odot$  \citep{Zhang2004,Vrsnak2007}. Unlike CME1, CME2 showed a gradual increase in the propagation and expansion acceleration, as shown in Figure (\ref{fig:kinematics}). We find CME1 as a fast speed with a rapid decrease in expansion followed by nearly constant expansion, while CME2 is a slow and gradually expanding CME. We note that the selected CMEs, especially CME2, could not be tracked at lower heights. Not being limb-CME for SOHO/LASCO and STEREO/COR, these events appeared beyond the coronagraphic occulter only after reaching a certain 3D radial height. Using the estimated kinematics, the thermodynamic behavior of CMEs associated with their expansion is explained in the following section.
\vspace{10pt}
\subsection{Implementing the FRIS model}

\begin{figure*}[ht]
        \includegraphics[scale=0.22,trim={4cm 1cm 20cm 2cm}]{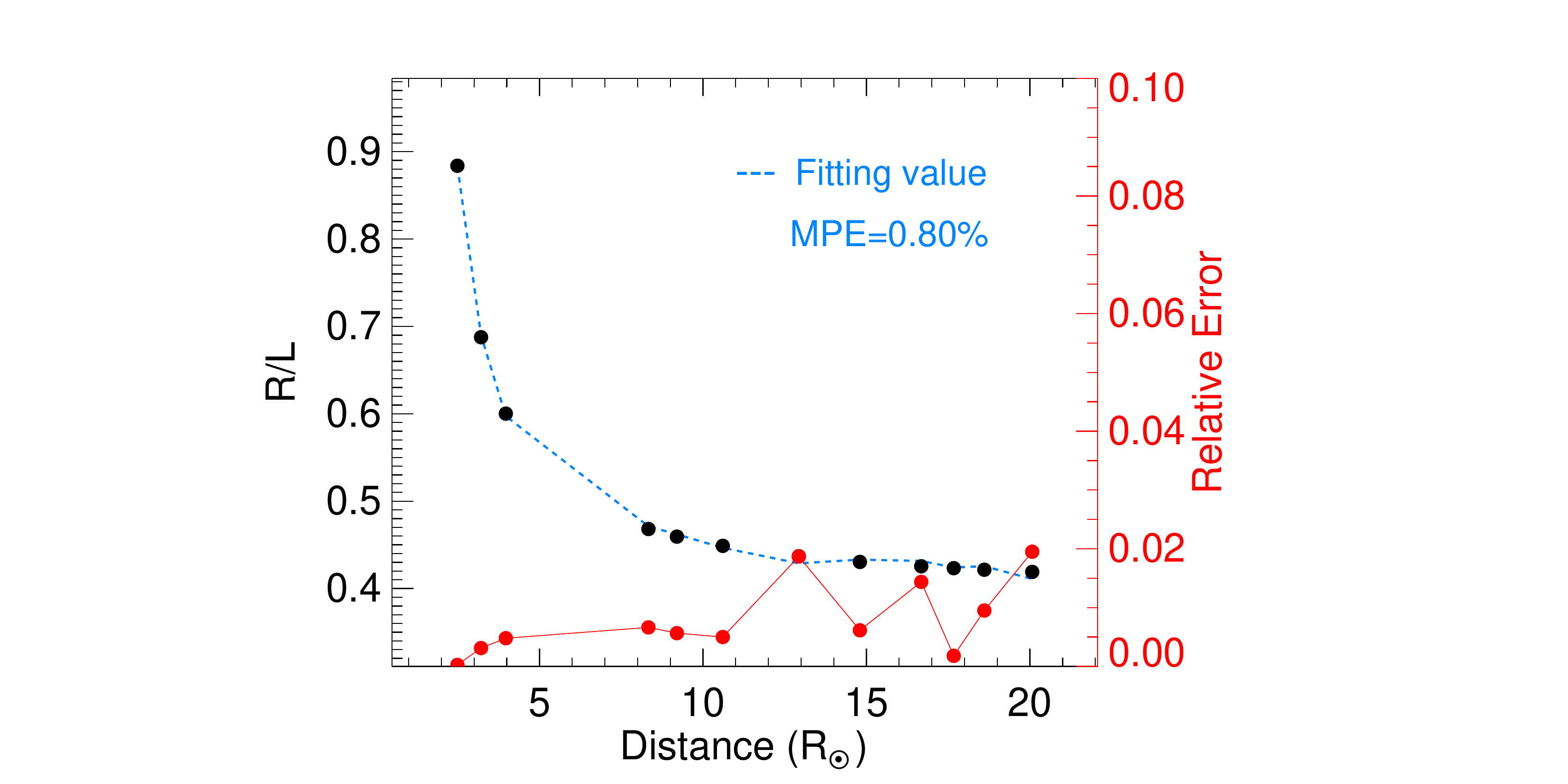}
        \includegraphics[scale=0.22,trim={0cm 1cm 22cm 2cm}]{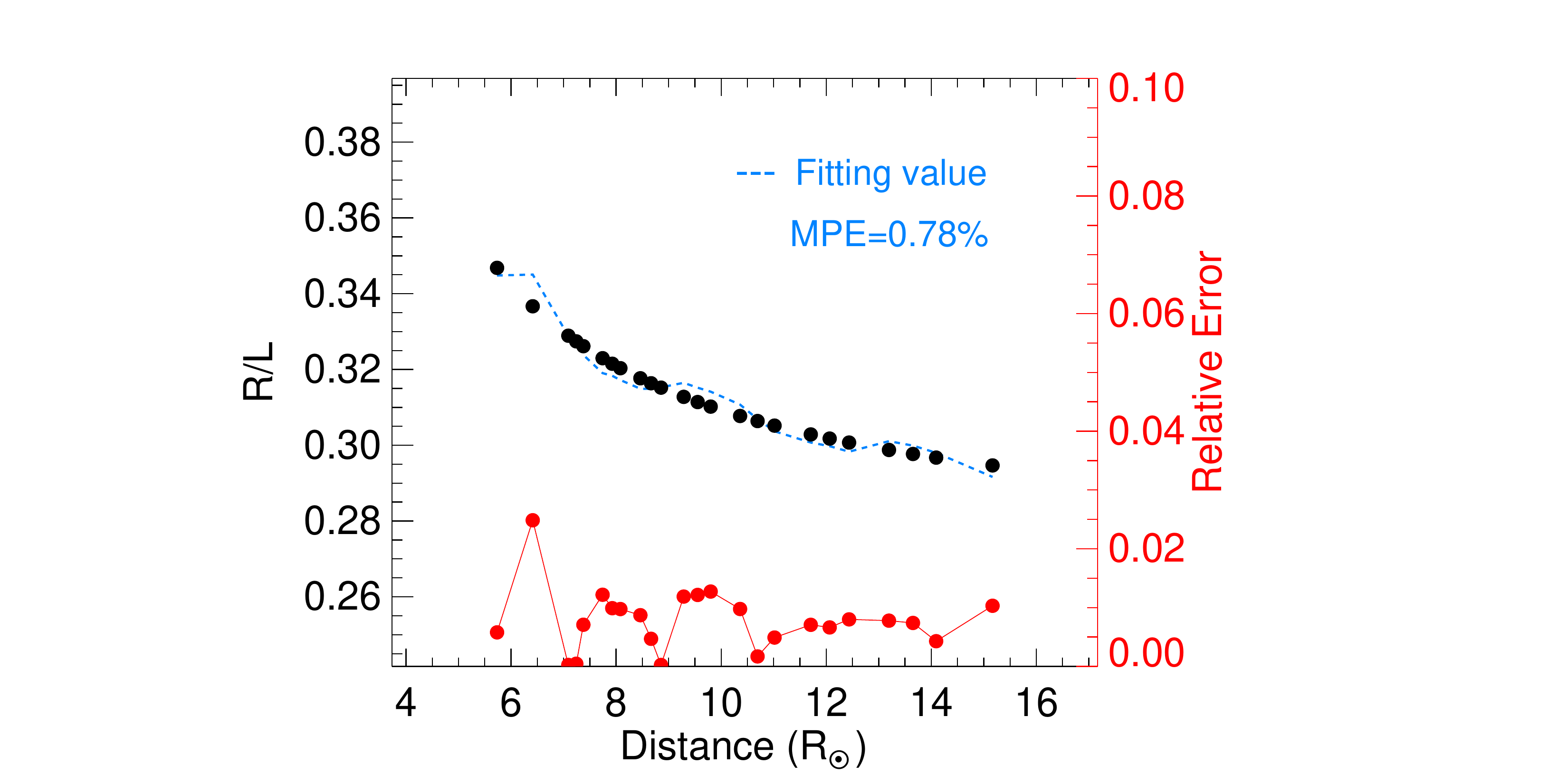}
        \caption{The profile of $R/L$ for CME1  and CME2 from the measurements (black; left-hand side of Equation \ref{eqn:fitting1}), the model-fitted result for this parameter (blue; right-hand side of Equation \ref{eqn:fitting1}), and the relative error (red) are shown. (MPE: Mean Percentage Error) }
        \label{fig:fitting}
\end{figure*}

As from the flow of the model and derived thermodynamic parameters, it is evident that the FRIS model requires the distance ($L$) of the centre of the CME flux-rope from the surface of the Sun, its propagation speed ($v_c$) which is the time-derivative of $L$, the radius of the flux-rope ($R$), its expansion speed ($v_e$) and their time derivatives. The distance $L$ can be expressed as $L= h-R-1R_\odot$ and $R$ can be estimated from GCS model derived aspect ratio ($\kappa$) and leading edge height ($h$) of CME flux-rope \citep{Thernisien2011}. To implement the FRIS model to the observations of CMEs, we have fitted the FRIS model derived Equation (\ref{eqn:fitting1}) by using the obtained 3D kinematic parameters as inputs, as shown in Figure (\ref{fig:fitting}). The fitting was done using the CURVE\_FIT routine in SCIPY, which enabled determining the values of five unknown coefficients $c_1$ to $c_5$ in the equation. Now, using the obtained kinematics of both CMEs and fitted coefficients, we estimate the internal forces and the thermodynamic properties of the selected CMEs (refer to Table \ref{tab:parameters}).

\subsubsection{Thermodynamic parameters}\label{sec:results_thermo}
FRIS model can estimate the several thermodynamic and plasma parameters of CMEs as shown in Table (\ref{tab:parameters}). In the present study, to keep the focus of the work, we determine the four most important parameters: the polytropic index ($\Gamma$), the average heating rate ($dQ/dt$), the average temperature ($\bar {T}$), and the average proton number density (${\bar{n}}_{p}$) of the CME flux-rope. For this purpose, we use the equations (\ref{eqn:lamb}), (\ref{eqn:heating}), (\ref{eqn:T}), and (\ref{eqn:np}) together with various 3D kinematic parameters of CMEs, and estimate the polytropic index ($\Gamma$), the average heating rate ($dQ/dt$), the average temperature ($\bar {T}$), the average proton number density ($ {\bar{n}}_{p}$), respectively. 

\begin{figure*}[ht]
         \includegraphics[scale=0.23,trim={4cm 1cm 20cm 0cm}]{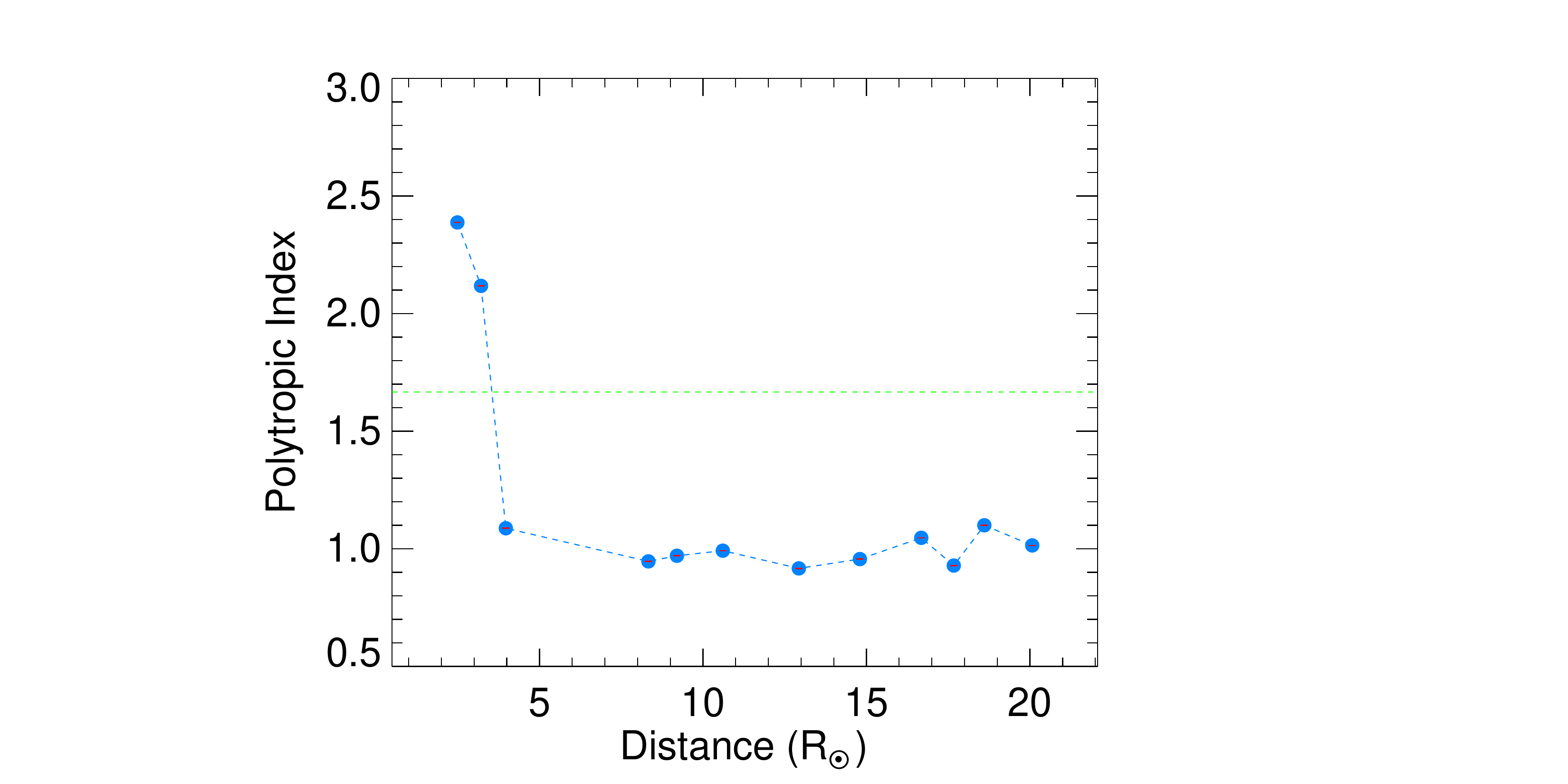}
         \includegraphics[scale=0.23,trim={0cm 1cm 22cm 2cm}]{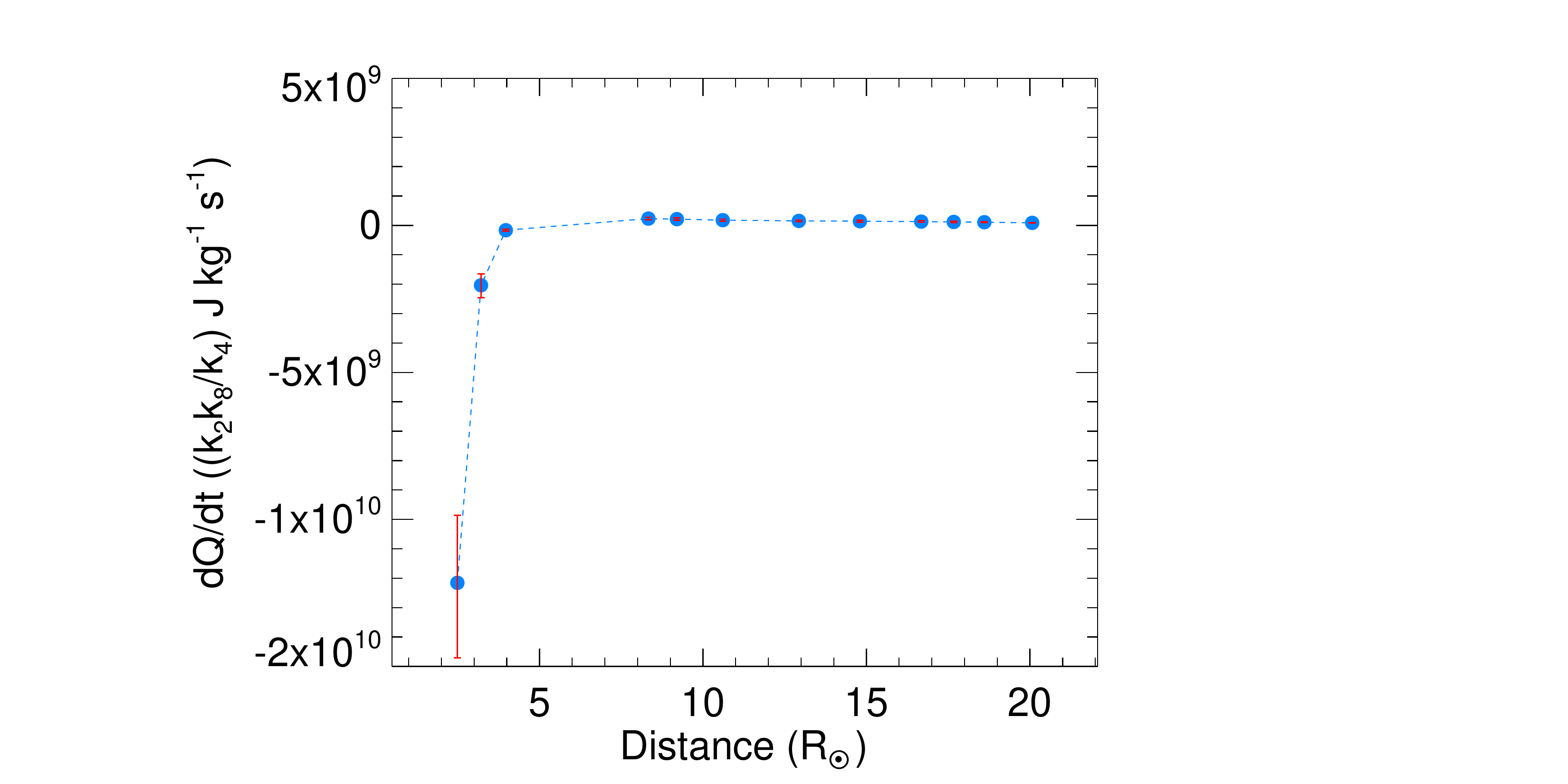}\\
         \includegraphics[scale=0.23,trim={4cm 1cm 20cm 2cm}]{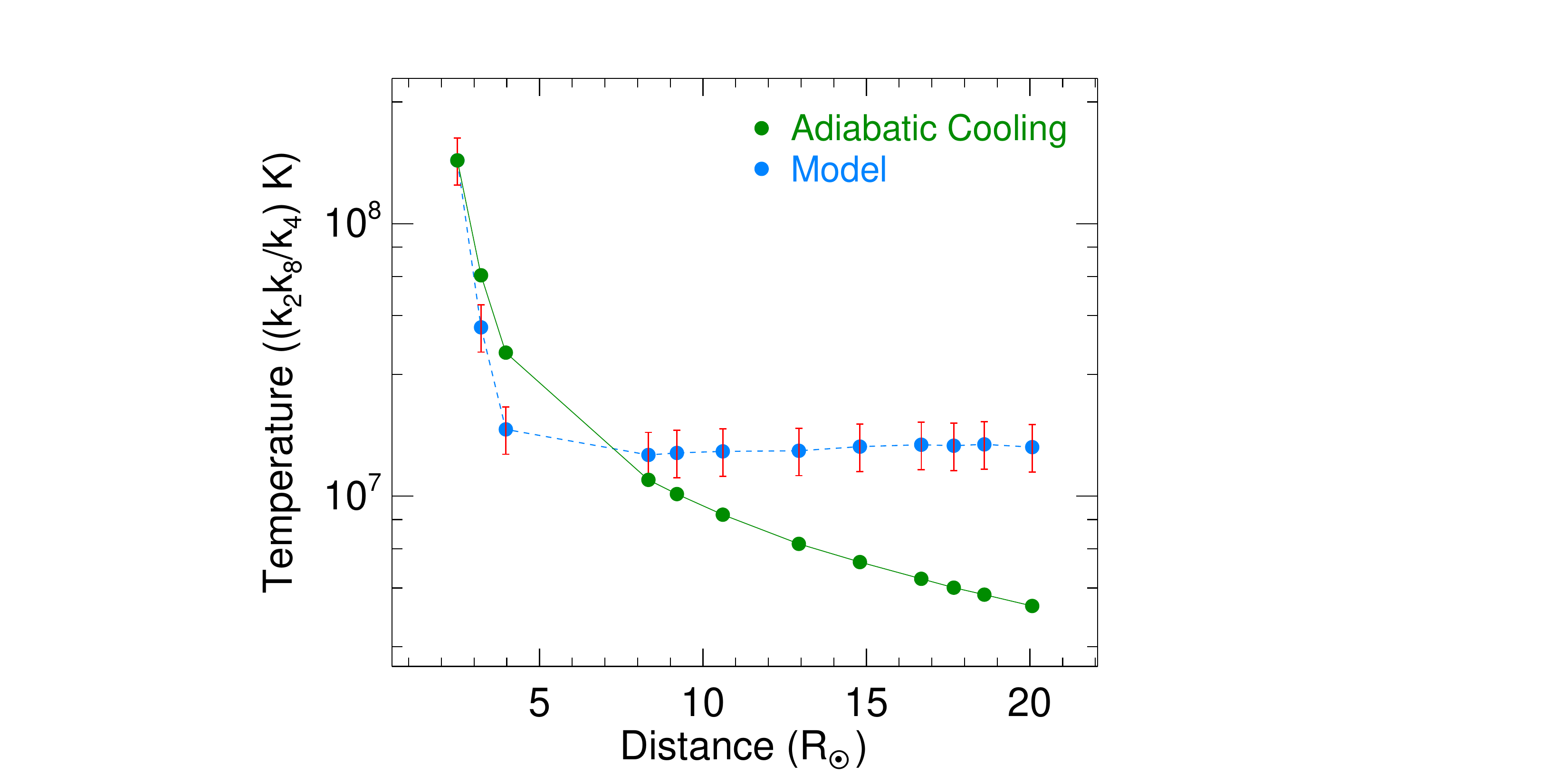}
         \includegraphics[scale=0.23,trim={0cm 1cm 22cm 2cm}]{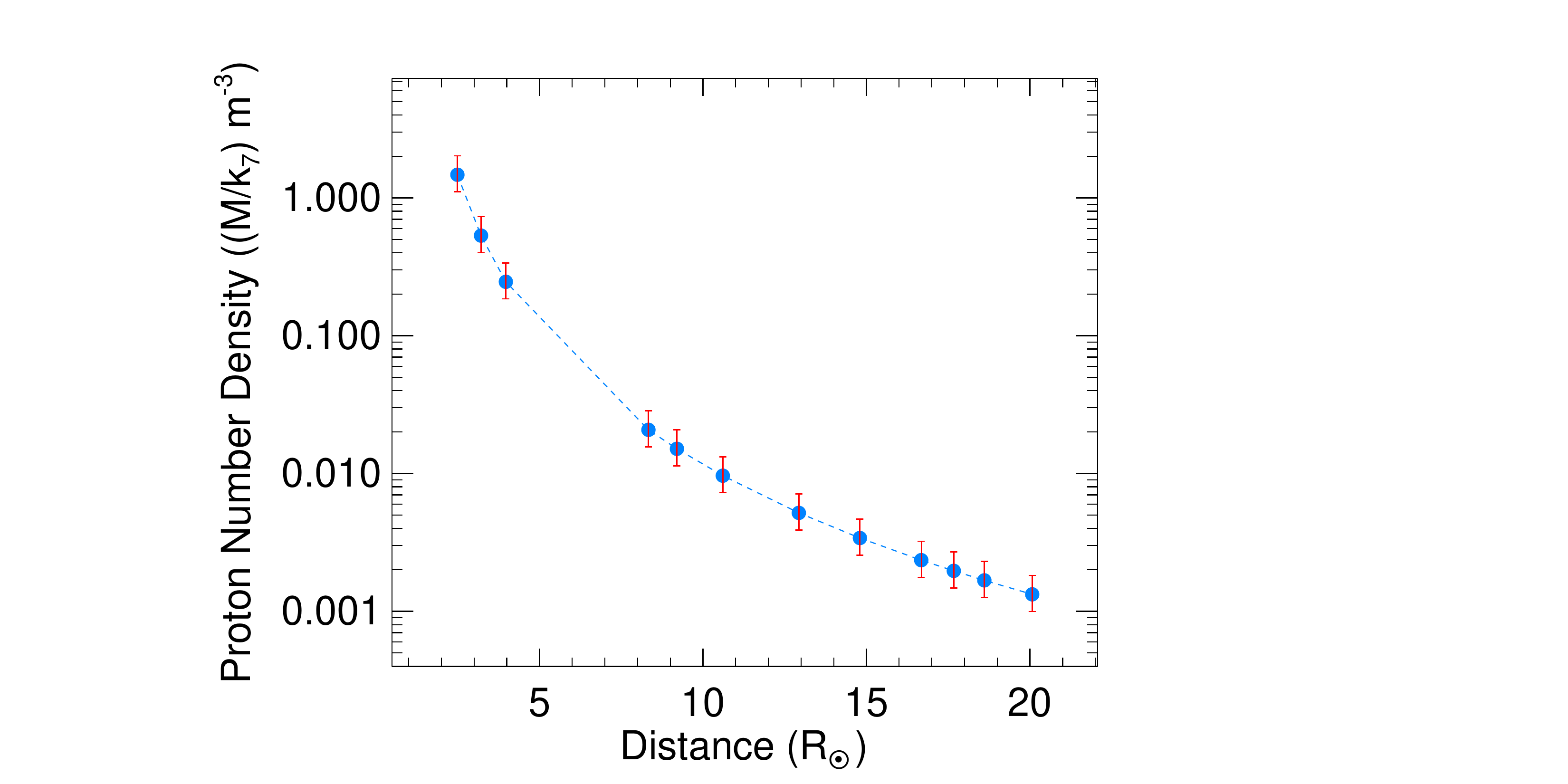}
         \caption{For the CME1: Variation of the polytropic index ($\Gamma$), average heating rate (dQ/dt), average temperature $(\bar{T})$, and average proton number density (${\bar{n}}_{p}$) of the CME with the heliocentric distance of the CME’s leading edge (h) is shown in the top-left, top-right, bottom-left and bottom right, respectively. The red vertical lines at each data point show the error bars derived by considering an error of 10{\%} in measurements of the flux-rope's leading edge height ($h$).}
        \label{fig:Thermo_para_CME1}
\end{figure*}

In this manner, implementing the FRIS model to CME1, we find that the value of $\Gamma$ is above the adiabatic index ($\Gamma = 5/3$) at the observed initial height from 2.5 $R_\odot$ to 3.2 $R_\odot$ (shown in top-left panel of Figure \ref{fig:Thermo_para_CME1}). It suggests the CME1 is releasing heat into the surrounding medium. Following this, the $\Gamma$ value rapidly decreases with the evolution of CMEs and reaches about 1.09 at around 4 R$_{\odot}$, beyond which the $\Gamma$ almost stays around the value one. It is evident that from 4.0 $R_\odot$, the value of $\Gamma < 5/3$ for CME1. The $\Gamma < 5/3$ value implies heat injections into the CME plasma, suggesting some physical processes responsible for heating the CMEs. It is evident that CME1 first releases heat coinciding with the fast decrease in acceleration of CME1 and afterward, during its gradual acceleration phase, absorbs heat, approaching almost an isothermal state ($\Gamma=1$) (top-left panel of Figure \ref{fig:Thermo_para_CME1}). The deviation from the adiabatic index value at higher heights indicates the CME is not thermally invariant.

We also estimated the average heating rate $(\bar \kappa=dQ/dt)$ of CME1 (top-right panel of Figure \ref{fig:Thermo_para_CME1}). The negative heating rate implies a heat-release into the surroundings, while the positive value means heat injection into the system. We find the heating rate to be a negative value since initially observed heights up to 3.2 $R_\odot$. Beyond this height, i.e., from 4 $R_\odot$, the heating rate maintains its positive value up to 20 $R_\odot$. However, we also note that the strength of heating $dQ/dt$ slightly decreases with increasing height after 8.4 $R_\odot$. This again confirms the finding from the polytropic index that CME1 is experiencing a continuous heat injection from 4 $R_\odot$. We also estimated the evolution of CME1 temperature from the FRIS model (bottom-left panel of Figure \ref{fig:Thermo_para_CME1}). We note that in the beginning, the temperature falls rapidly with a distance up to 4 $R_\odot$, following which the temperature increases slowly as the CME1 propagates away from the Sun. These results confirm the initial heat-release followed by a positive heating of CME1. The estimates of average proton density evolution of CME1 shown in the bottom-right panel of Figure (\ref{fig:multi-wavelength_CME1}), suggest that the density (${\bar {n}}_{p}$) decreases faster with height at up to 4 $R_\odot$ during the fast decrease in the acceleration phase of CME1. Subsequently, ${\bar {n}}_{p}$ decreases slowly compared to the first phase of CME1's propagation. The faster-decreasing density before 4 $ R_\odot$ suggests that the CME1 went through a rapid expansion at the beginning of its journey. The overall decrease in density refers to the expansion of the CME flux-rope. Interestingly, despite a positive heating rate and a $\Gamma$ value of less than $5/3$ from 4 $ R_\odot$, the decrease in temperature continues up to a height 20 $R_\odot$. It is possible that between 3.2 to 4 $R_\odot$ heights, the heat added into the system is almost sufficient for compensating for the faster cooling due to the rapid expansion of CME1. In fact, the heat added inside CME1 beyond 8.4 $ R_\odot$ could increase its temperature as the CME continues its gradual or nearly constant expansion phase.

\begin{figure*}[ht]    
        \includegraphics[scale=0.23,trim={4cm 1cm 20cm 0cm}]{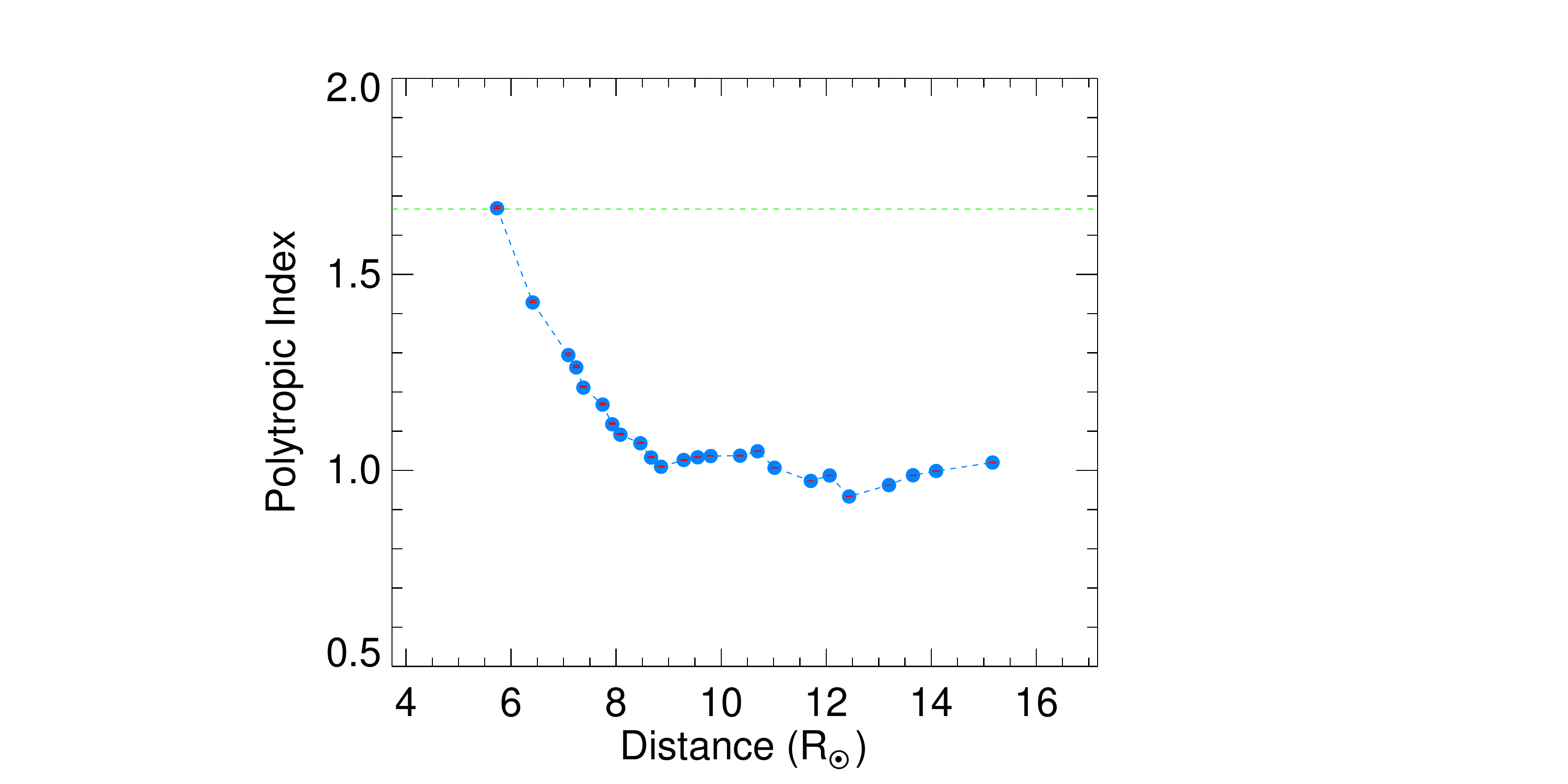}       
         \includegraphics[scale=0.23,trim={0cm 1cm 22cm 0cm}]{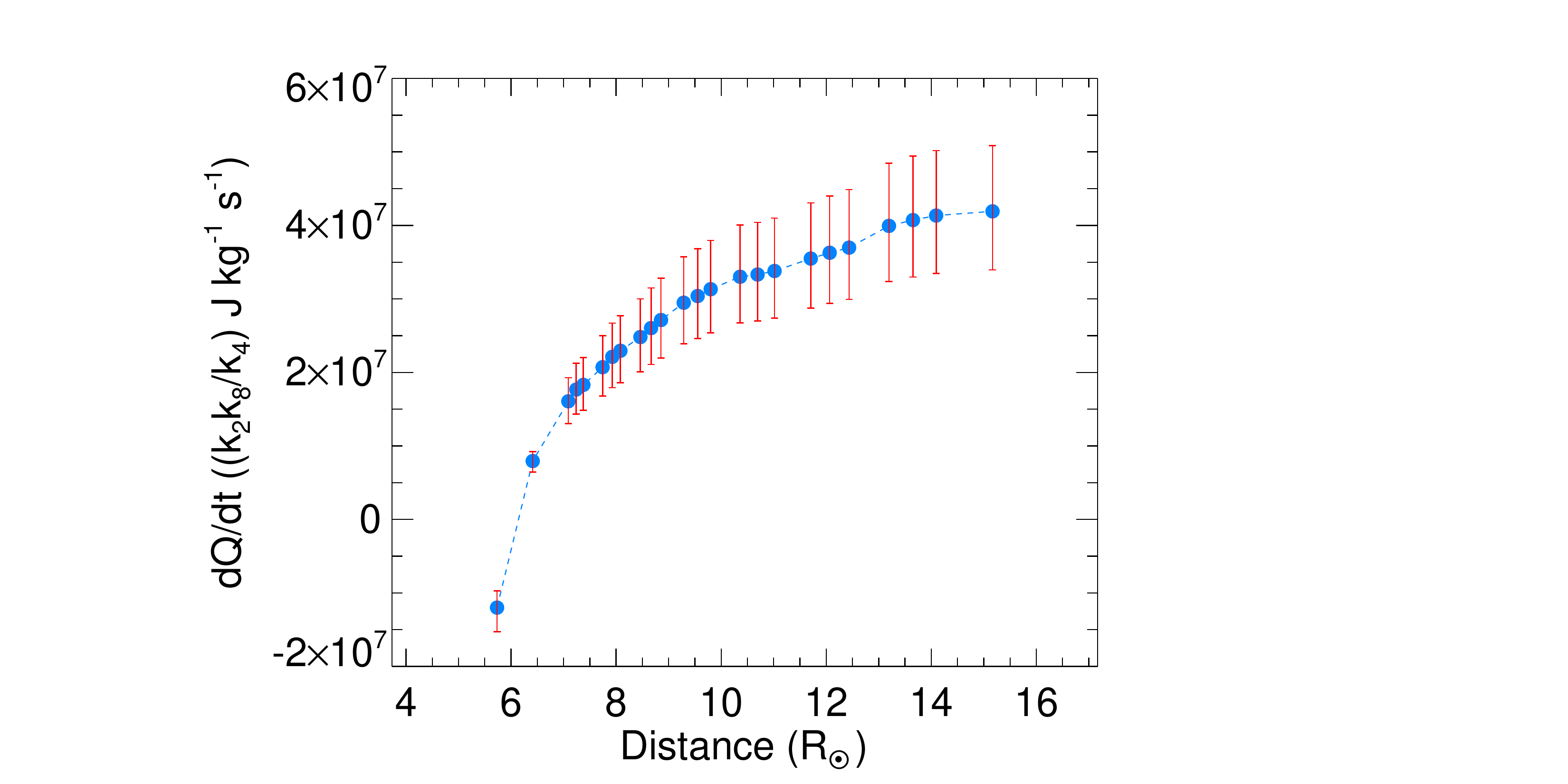}\\
         \includegraphics[scale=0.23,trim={4cm 1cm 20cm 2cm}]{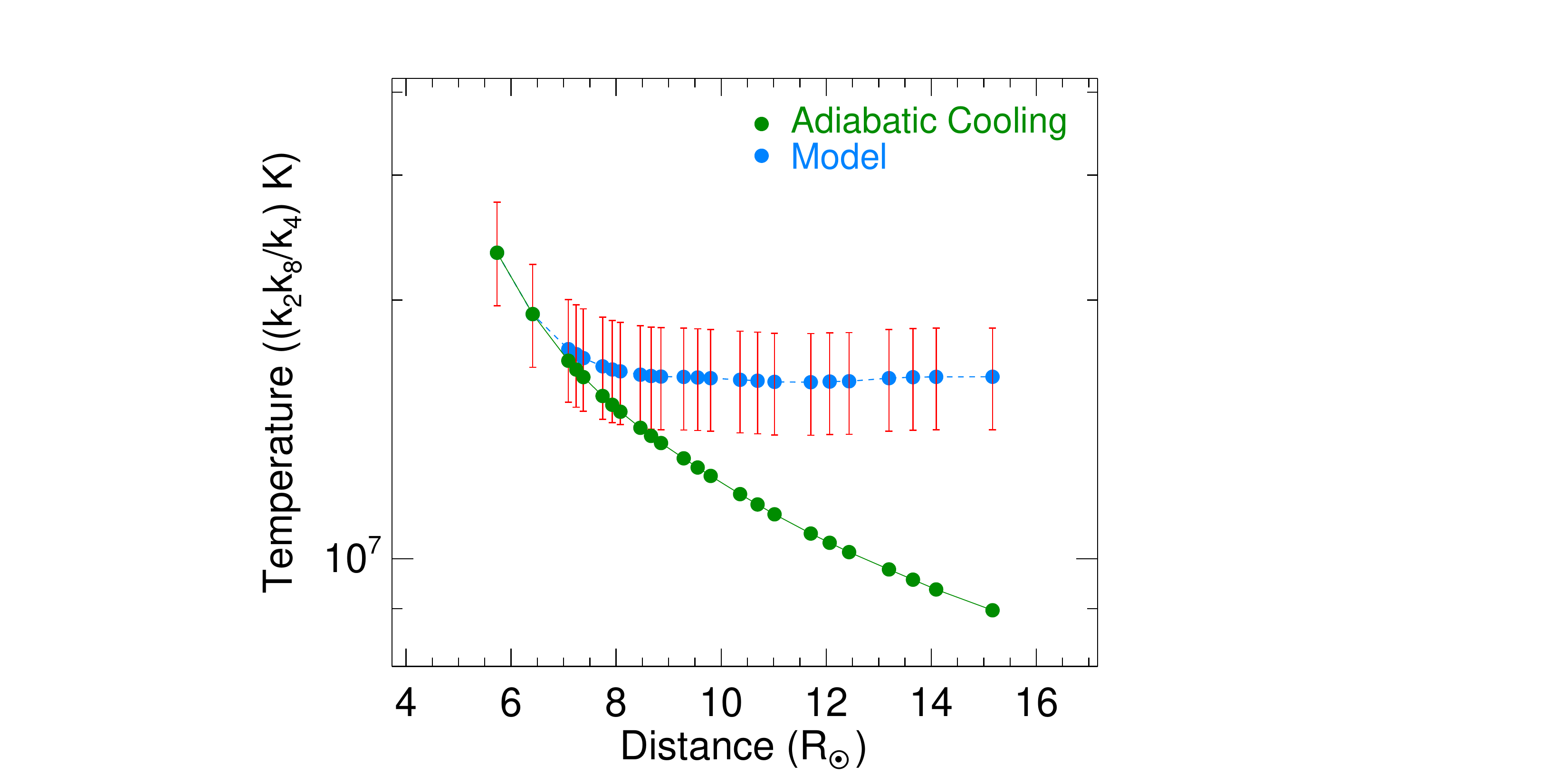}
         \includegraphics[scale=0.23,trim={0cm 1cm 22cm 2cm}]{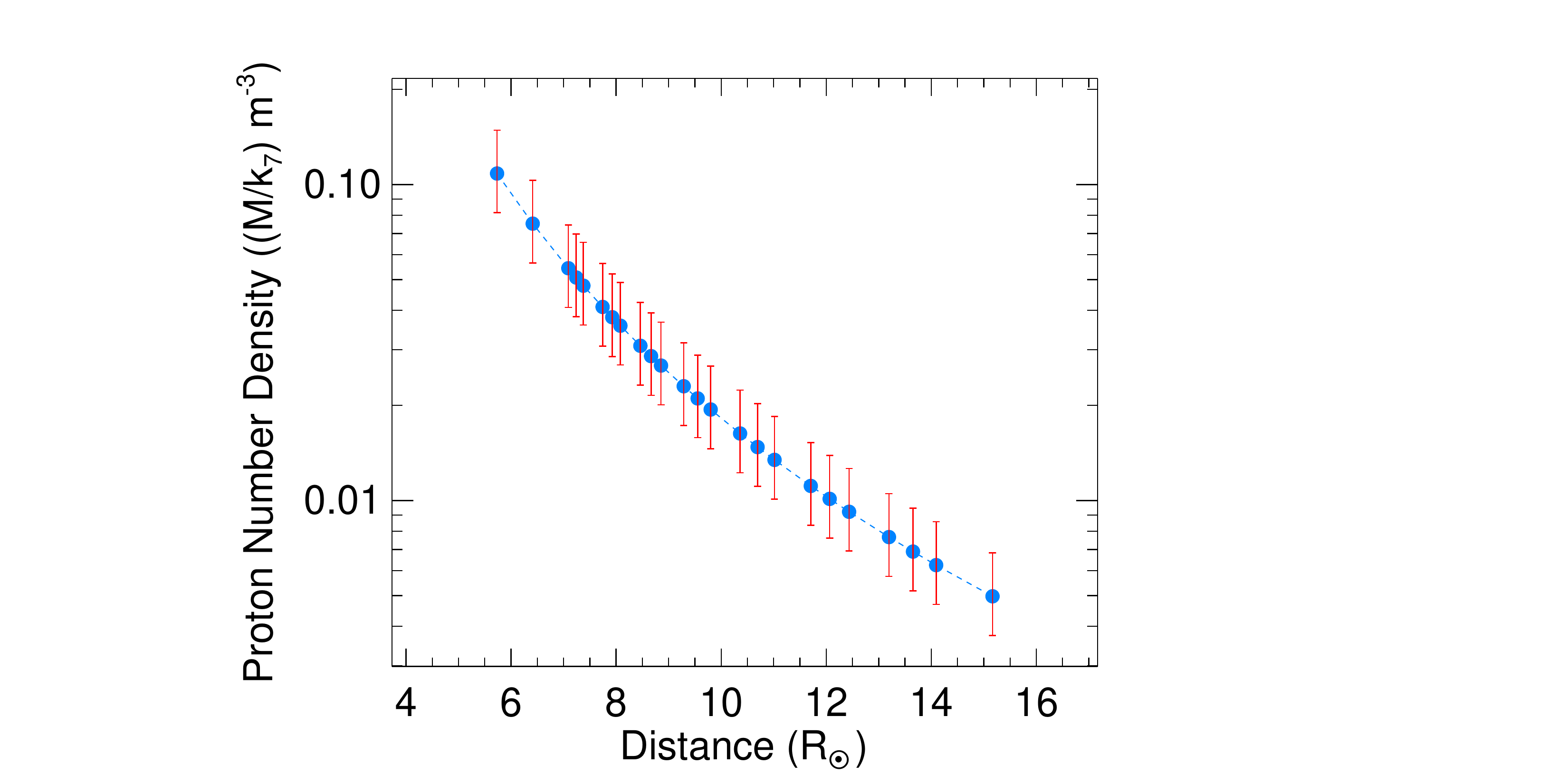}
         \caption{For the CME2: Variation of the Polytropic index ($\Gamma$), average heating rate (dQ/dt), average temperature $(\bar{T})$, and average proton number density (${\bar{n}}_{p}$) of the CME with the heliocentric distance of the CME’s leading edge (h) is shown in the top-left, top-right, bottom-left and bottom right, respectively. The red vertical lines at each data point show the error bars derived by considering an error of 10{\%} in measurements of the flux-rope's leading edge height ($h$).}
        \label{fig:Thermo_para_CME2}
\end{figure*}

We also implemented the FRIS model to CME2 observations, and the derived thermodynamic parameters for the CME2 are shown in Figure (\ref{fig:Thermo_para_CME2}). Except for the initial height at 5.6 $R_\odot$, the polytropic index $\Gamma$ for CME2 is below the adiabatic index value (top-left panel of Figure \ref{fig:Thermo_para_CME2}), which suggests a continuous injection of heat into the CME plasma. The $\Gamma$ value is closer to the isothermal index ($\Gamma=1$) after 8 $R_\odot$, which means the heat added into the CME2 can maintain its near isothermal state despite its increasingly larger expansion. The estimated heating rate for CME2, shown in the top-right of Figure (\ref{fig:Thermo_para_CME2}), increases with height as the CME flux-rope propagates up to 15 $R_\odot$. The positive value of $dQ/dt$ suggests an absorption of heat, while its increasing trend indicates increasingly larger heat added to the system. The derived temperature profile for CME2 shows a decrease in its value from the initially observed height up to 8 $R_\odot$ (bottom-left of Figure \ref{fig:Thermo_para_CME2}), after which it maintains a constant temperature for the remaining observed heights. It is possible that increasingly larger heat added to the CME2 nearly balances its expected cooling due to the expansion. The average proton number density (${\bar {n}}_{p}$) for the CME2 is found to decrease gradually with distance (bottom-right of Figure \ref{fig:Thermo_para_CME2}).

In our study, we also compared the FRIS model-derived temperature with the adiabatic temperature profile, which assumes no heat-release or absorption by CMEs. For both the CMEs, the adiabatic temperature profile is obtained using the equation $ \frac{T_{i+1}}{T_i}=\left(\frac{p_{i+1}}{p_i}\right)^{1-\frac{1}{\gamma}}$; where $T_i$ and $T_{i+1}$ refers to the temperature at two evolving times/distances (data points), similarly for $p_i$ and $p_{i+1}$. To estimate the adiabatic temperature profile of CMEs, we use FRIS model-derived pressure at each time step while FRIS model-derived temperature is used for only the first observed time (data point). In the absence of additional heating/cooling, the adiabatic temperature (i.e., cooling) should be consistent with the observed expansion of CMEs and consequent decrease in plasma pressure. Obviously, for our selected CMEs experiencing heat-release (or heat-absorption), the FRIS model-derived temperature will be smaller (or larger) than the adiabatic temperature. We found the model-derived temperature for the CME1 is smaller than the adiabatic temperature at the beginning up to 4 $R_\odot$. As the CME1 absorbs sufficient heat, its temperature slowly rises after 8.4 $R_\odot$, and the adiabatic temperature falls below the model-derived temperature value (bottom-left panel of Figure \ref{fig:Thermo_para_CME1}). The model-derived temperature for CME2 starts to fall almost similar to the adiabatic temperature, but as the CME2 experience significant heating, its temperature stays above the adiabatic value between 7 to 15 $R_\odot$ (bottom-left panel of Figure \ref{fig:Thermo_para_CME2}). This result suggests, firstly, there is a transfer of heat from/to the CME during the heliospheric propagation of the CMEs; secondly, the CME temperature does not fall as per the observed expansion of CME, considering it to be in the adiabatic state.

\subsubsection{Internal Forces}\label{sec:results_forces}

\begin{figure*}[ht]
\centering
        \includegraphics[scale=0.24,trim={18cm 1cm 10cm 0cm}]{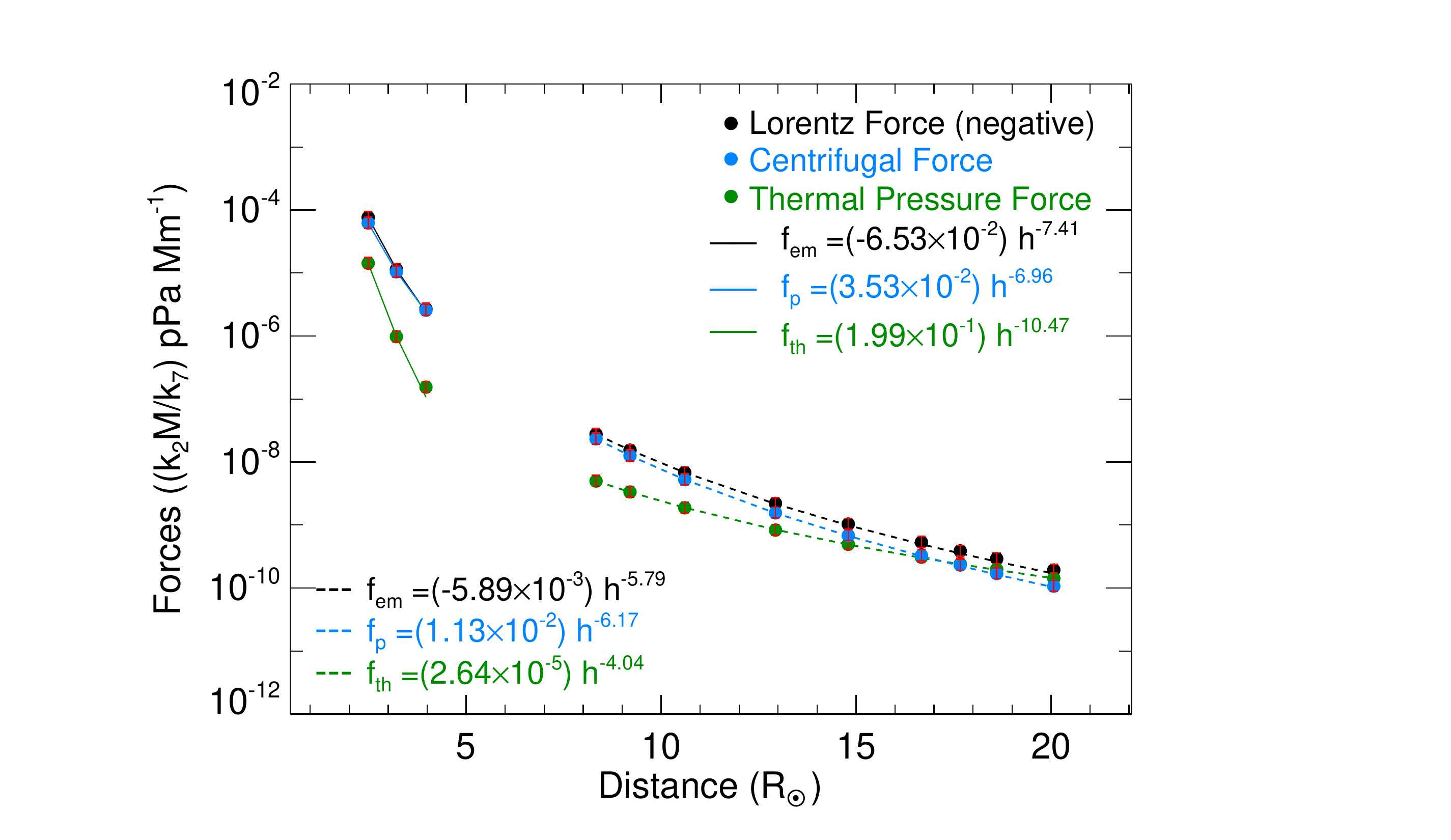}
         \includegraphics[scale=0.24,trim={2cm 1cm 20cm 0cm}]{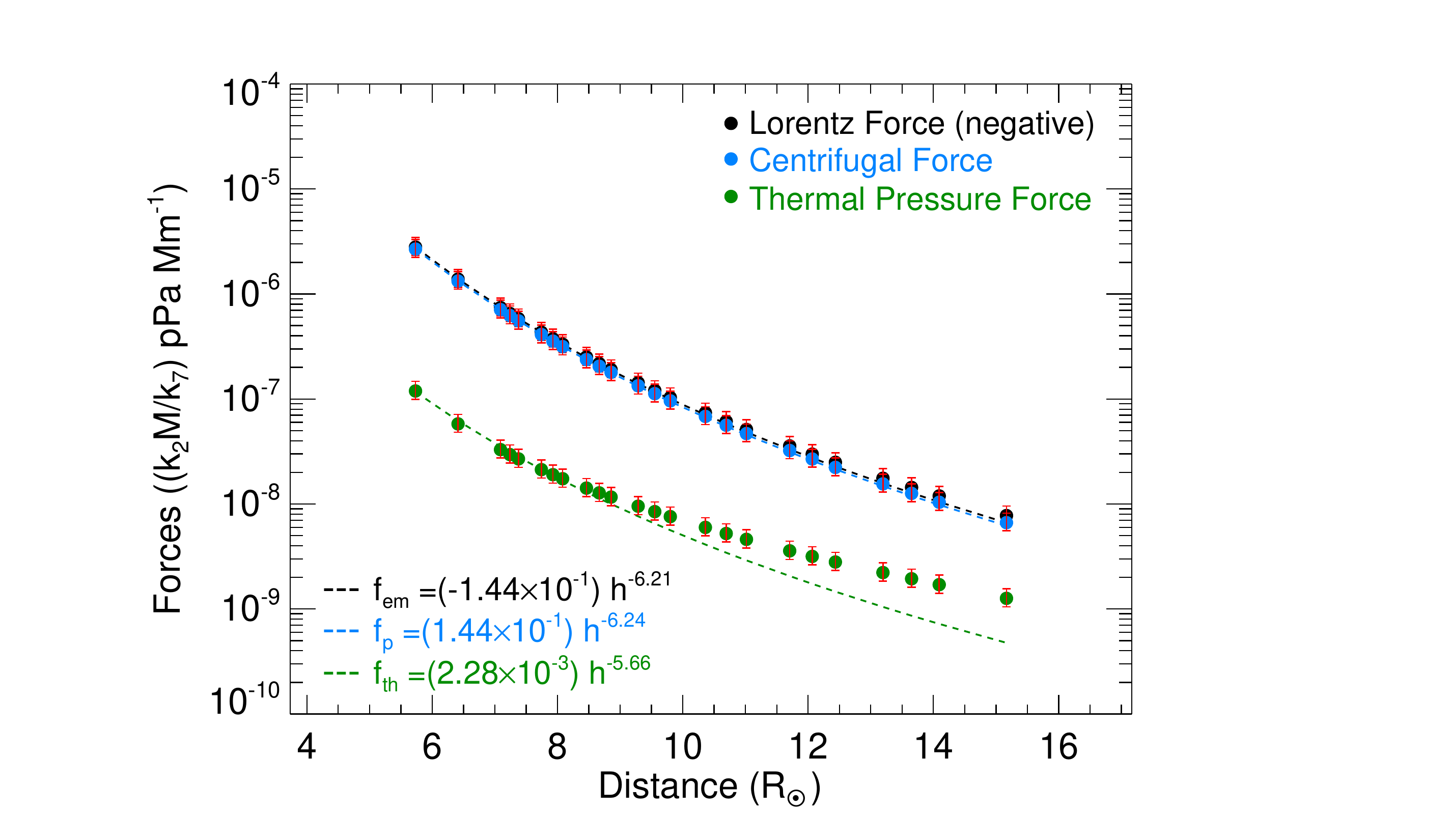}
        \caption{The FRIS model-derived average internal forces, such as Lorentz force, Thermal pressure force, and Centrifugal force, that is responsible for the radial expansion of the flux-rope of CME1 (left) and CME2 (right). The red vertical lines at each data point show the error bars derived by considering an error of 10{\%} in measurements of the leading edge height of the flux-rope ($h$). The solid and dash lines show the power-law fitted values for the model-derived internal forces.}
        \label{fig:internal_forces}
\end{figure*}

In addition to thermodynamic parameters, the FRIS model can also derive the internal forces such as average Lorentz force (${\bar f}_{em}$), average thermal pressure force (${\bar f}_{th}$) and average centrifugal force (${\bar f}_{p}$) using equations (\ref{eqn:fem_final}), (\ref{eqn:fth_final}), and (\ref{eqn:fp_final}), respectively. For the selected CME1 and CME2, we estimated the evolution of these three forces acting on the CMEs during their propagation away from the Sun, as shown in Figure (\ref{fig:internal_forces}). These forces govern the internal dynamical processes responsible for the observed radial expansion profile of the CMEs. We have found the direction of the Lorentz force is negative, while the thermal pressure and centrifugal force are positive for both the CMEs during their complete journey, as observed. This means that the Lorentz force is acting toward the centre of the flux-rope while the other two forces, the thermal pressure force and the centrifugal force, are acting away from the centre of the flux-rope. Thus, our findings suggest that ${\bar f}_{em}$ is preventing the expansion while ${\bar f}_{th}$ and ${\bar f}_{p}$ are responsible for the expansion of the flux-rope. This is true irrespective of CMEs with fast or slow speed profiles.

Further, we found all three forces decrease faster in CME1 at the beginning (up to 4 $R_\odot$) followed by a gradual decrease up to 20 $R_\odot$. However, the decrease rate for forces is moderate for CME2 during its observed journey (5-15 $R_\odot$). We fitted a power law to the measured internal forces to quantify their decreasing rate for both selected CMEs. The fitted power law function is $y=m.h^{-c}$; where $m$ and $c$ are fitted coefficients, and $h$ is the leading edge height of the CME flux-rope. We note that selected CME1 shows two-phase kinematics, i.e., an initial rapid decrease in acceleration followed by a constant acceleration phase. This gives rise to two-phase evolution in the FRIS model-derived internal force parameters for CME1, which is fitted using two power-law profiles. For the initial-phase fitting of CME1, where it shows a rapid decrease in acceleration, we find that ${\bar f}_{th}$ is decreasing with the fastest rate as ${\bar f}_{th} \propto h^{-10.47}$ whereas ${\bar f}_{em}$ and ${\bar f}_{p}$ is decreasing with slower rate as ${\bar f}_{em} \propto h^{-7.41}$ and ${\bar f}_{p}\propto h^{-6.96}$. The power law fitting for the second-phase of CME1 shows the decrease in ${\bar f}_{em}$, ${\bar f}_{p}$ and ${\bar f}_{th}$ is proportional to $h^{-5.79}$, $h^{-6.17}$, and $h^{-4.04}$, respectively (left panel of Figure \ref{fig:internal_forces}). The CME2 was fitted with a single power law profile, as shown in the right panel of Figure (\ref{fig:internal_forces}). The decreasing in ${\bar f}_{em}$, ${\bar f}_{p}$ and ${\bar f}_{th}$ is proportional to $h^{-6.21}$, $h^{-6.24}$, $h^{-5.66}$, respectively. Thus for CME2, the decreasing rate is highest for the ${\bar f}_{p}$ and is lowest for the ${\bar f}_{th}$. However, we have noticed that the ${\bar f}_{th}$decreases more slowly at higher heights after around 9 $R_\odot$. A critical thing to notice is that if a CME is slow and gradually accelerating, all the internal forces decrease gradually. In contrast, the internal forces decrease much faster for a CME with fast and rapidly decreasing acceleration, as in the initial phase of CME1. Also, we have found that a CME can have different rates of change in forces governing its internal dynamics for different segments of its journey. Thus, we can infer one interesting result: the internal dynamics behave similarly and may govern the CME's global kinematic profile.

In each CME1 and CME2, we find that ${\bar f}_{em}$and ${\bar f}_{p}$ have a similar magnitude at the beginning but ${\bar f}_{p}$ drops faster than ${\bar f}_{em}$ as CMEs propagate outward. For CME1, the ${\bar f}_{th}$ always has a lower magnitude than the other two forces but decreases rapidly compared to the other two at initial heights up to 4 $R_\odot$. Thus, the large magnitude of the radially inward force, ${\bar f}_{em}$, and its slight decrease is responsible for the initial decrease in expansion acceleration. However, the ${\bar f}_{th}$ has a much lower magnitude than the other two forces for CME2; hence the net force ${\bar f}_{net}={\bar f}_{em}+{\bar f}_{th}+{\bar f}_{p}$ is negative up to 6.6 $R_\odot$ and leads to a negative expansion acceleration. However, the sum of positive forces, i.e., ${\bar f}_{p}$ and ${\bar f}_{th}$ overtaking the negative ${\bar f}_{em}$ will allow the CME to gain a positive expansion acceleration. In both observed cases, we also note that the contribution of ${\bar f}_{p}$ is always more than ${\bar f}_{th}$ at initial heights. This may be reversed at some higher heights as indicated by their rate of decrease. In case of CME1, the ${\bar f}_{th}$ overtakes ${\bar f}_{p}$ and has higher magnitude beyond 18 $R_\odot$. From the trends of decrease of forces, it is obvious that at some higher heights (farther out of final heights estimated for selected CMEs), ${\bar f}_{th}$ will have a higher magnitude than the other two forces for the selected CMEs. Thus we can infer that ${\bar f}_{th}$ can be primarily responsible for the expansion of the CME at heights much away from the Sun. It is imperative to track CMEs up to much higher heights in the interplanetary medium and examine their thermodynamics, kinetics and different forces.

\subsection{Multi-wavelength Imaging Observations of the source region of CMEs}

Examining the thermodynamic state of a CME flux-rope at its source region is also important. This can help to understand the inherent thermal state of CMEs, which can further be associated with their evolving thermal states at higher coronal heights. As we already described the FRIS model-derived estimates of thermodynamic parameters at heights where the kinematics of CMEs could be determined unambiguously using coronagraphic observations, now we analyse the flux-rope thermal state using multi-wavelength observations at their birth in the lower corona. For this purpose, we use high-resolution EUV imaging observation of \textit{Atmospheric Imaging Assembly (AIA)}, \textit{Heliospheric Magnetic Imager (HMI)} onboard \textit{Solar Dynamics Observatory (SDO)}, STEREO, SOHO and  \textit{Global Oscillation Network Group (GONG)}/H$_{\alpha}$. 

\begin{figure*}[ht]
  \includegraphics[trim = 0.0cm 0.0cm 1.0cm 0.0cm, scale=1.0]{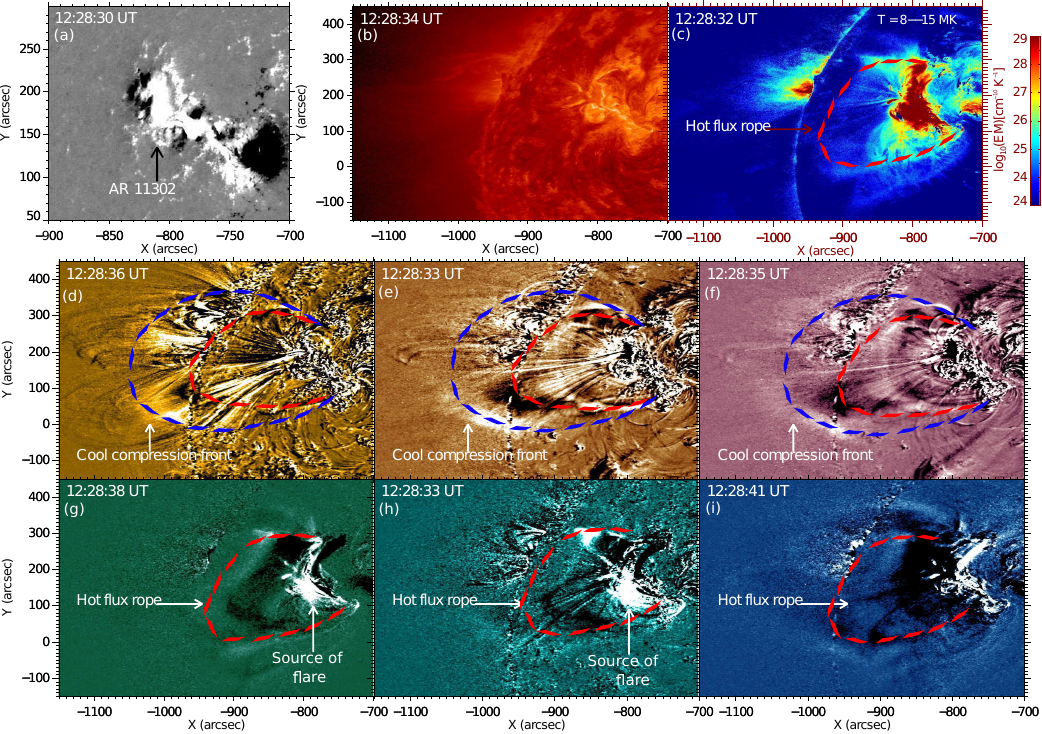}
row\caption{Multi-wavelength view of magnetic flux-rope as seen in different wavebands of SDO/AIA. Top row (panels a--c): The hot flux-rope formed above an active region, AR 11302, as shown by SDO/HMI line-of-sight magnetograms (panel a). The SDO/AIA 304 {\AA} image depicts the absence of flux-rope/eruption (panel b). The panel (c) indicates the presence of a hot flux-rope (8--15 MK) in the DEM map.  Middle row (panels d--f): It shows the distinct feature of the magnetic flux-rope in its eruptive phase. The cyan dotted lines show the cool compression front, followed by the dark cavity region of the flux-rope, which best appears at coronal temperature (0.6--2 MK) in AIA 171 {\AA}, 193 {\AA}, and 211 {\AA} wavebands.  Bottom row (panels g--i): The hot channel of the flux-rope has been shown via dotted red lines in the hotter wavebands of AIA (2.5--10 MK, i.e., 94 {\AA}, 131 {\AA}, and 335 {\AA}). It should be noted that cool coronal images (middle panel) show the very faint signature of this hot channel of the flux-rope.}
 \label{fig:multi-wavelength_CME1}
\end{figure*}

\subsubsection{CME1: CME of 2011 September 24}

We have used the HMI magnetogram to identify the source region of this eruption, and it is associated with a $\beta\gamma$-type sunspot in AR 11302 at NE limb, which possesses several M-class and one X-class flare within 24 hours (panel a of Figure \ref{fig:multi-wavelength_CME1}). The build-up energy process, formation, evolution, and eruption of the flux-rope, as observed, indicate that the CME1 flux-rope is hot \citep{Zhang2012}, which is associated with intense solar flare as shown in panels g and h of Figure (\ref{fig:multi-wavelength_CME1}). On 2011 September 24, at $\approx$12:33 UT, we found evidence of an M7.1 class flare in AR 11302 that also led to an outburst of plasma that appeared as a fast halo CME, i.e., CME1 in the LASCO field-of-view. { It should be noted that the initiation of M-class flare start from 12:33 UT. We have indicated the source region of the initiation of flare (enhancement in the brightening in hot wavebands of AIA EUV channels i.e., 94 and 131 {\AA}; panels g and h of Figure~\ref{fig:multi-wavelength_CME1})). Before the initiation of the flare, the hot flux rope destabilize and started to lift-up (panels d--i, Figure~\ref{fig:multi-wavelength_CME1})}. The observations of the CME1 flux-rope show three components as a leading compressed bright front, followed by a dark cavity region and hot channel of the flux-rope \citep{Zhang2012}. The multi-wavelength view of the eruptive flux-rope and its association with the hot channel is shown in the middle and bottom panel (d--i) of Figure (\ref{fig:multi-wavelength_CME1}). This eruptive flux-rope is diffuse; therefore, we use base difference images of SDO/AIA filters to improve the contrast concerning the back coronal region. The appearance of the hot channel of the flux-rope started to lift $\approx$12:05 UT, which was about 20 minutes earlier than the initiation of the flare. The hot channel of the flux-rope (manually tracked by red dotted lines) appears only in the hot waveband of the AIA filters (e.g., 94 {\AA} (6.3 MK), 131 {\AA} (10 MK), and 335 {\AA} ({2.5 MK}), panels g, h, and i of Figure \ref{fig:multi-wavelength_CME1}); however, the corresponding cool component of the flux-rope is absent (e.g., 171 {\AA}, 193 {\AA}, and 211 {\AA}; panels d, e, and f of Figure \ref{fig:multi-wavelength_CME1}). We note that the flux-rope seen in the hot channel pushes the overlying field lines, and the flux-rope's bright but cooler front is best visible at coronal temperature (171 {\AA} (0.6 MK), 193 {\AA} (1.2 MK), 211 {\AA} (2 MK); panels d, e, and f of Figure \ref{fig:multi-wavelength_CME1}). This cool component is considered a compression front or leading edge (LE) of the flux-rope that appears due to the push of hot flux-rope in the surrounding coronal magnetized plasma (manually tracked blue dotted lines; panels d, e, and f of Figure \ref{fig:multi-wavelength_CME1}). Such scenarios support the onset of torus instability to trigger the flux-rope \citep{Kliem2006, Zhang2012}. The reconnecting current sheet might be formed between the overlying cooler bright front and the hot channel of the flux-rope, which can partly be responsible for further heating the flux-rope. We have performed the differential emission measure (DEM)\citep{Cheung2015} to understand the thermal behavior of the eruptive flux-rope. We found that the hot channel of the flux-rope has a wide range of temperatures and is as high as $\approx$ 8 to 15 MK temperature (panel c of Figure (\ref{fig:multi-wavelength_CME1}); {red} dotted lines). The temperature of the compressed cooler front is possibly spread around 1 to 3 MK since it is visible at 211 Å, 193 Å, and 171 Å. This indicates that the CME1 flux-rope is heated at its birth itself, which is expected to release heat during its expanding propagation in the outer corona. This is consistent with the FRIS model-derived estimates of thermodynamic parameters of CME1, which shows that a decrease in temperature continues to 4 $R_\odot$, and also, the polytropic index remains above the adiabatic index.

\begin{figure}[ht]
  \includegraphics[trim = 0.0cm 0.0cm 1.0cm 0.0cm, scale=1.0]{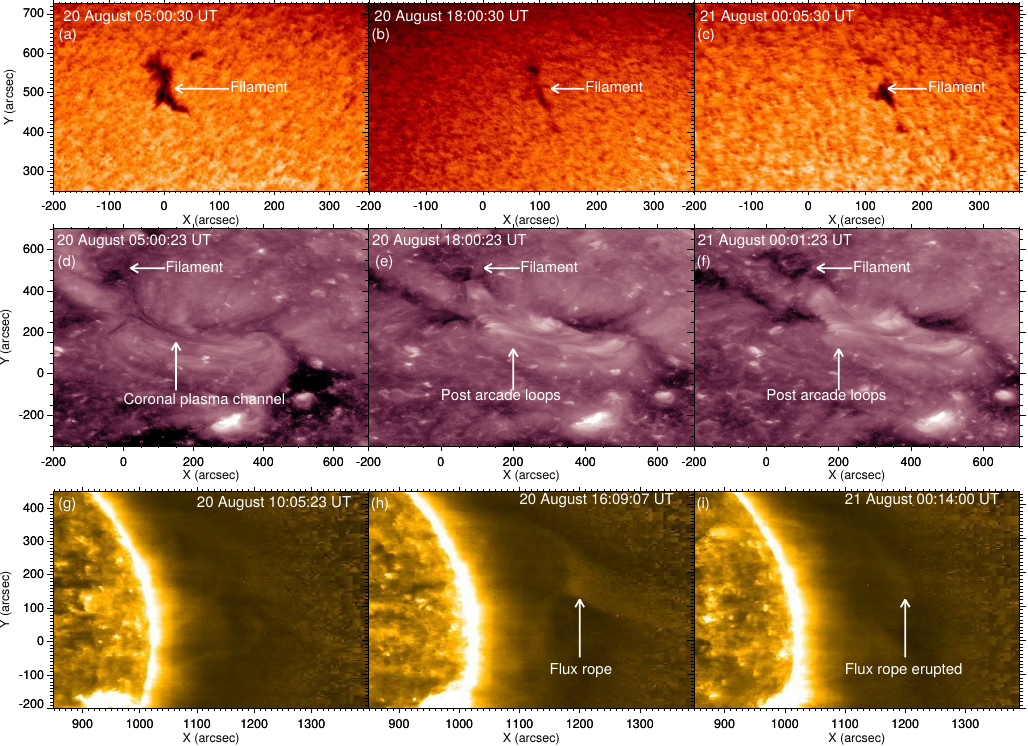}
  \caption{Top row (panels a--c): The sequence of images of GONG/H$_{\alpha}$ shows the gradual eruption of a quiescent filament on 2018 August 20. It is associated with an elongated filament channel containing no cool component but a hot plasma channel. Middle row (panels d--f): Temporal evolution of coronal plasma channel (hot filament channel) lying between two coronal holes as observed by SDO/AIA 211 {\AA}.  Bottom row (panels g--i): The formation of flux-rope and its eruption lying over the coronal plasma channel is shown in STEREO-A/EUVI 171 {\AA} on 2018 August 20.}
  \label{fig:multi-wavelength_CME2}
  \end{figure}

\subsubsection{CME2: CME of 2018 August 20}

To identify the source region of the selected CME2 that appeared as a slow and faint CME in the LASCO field-of-view on 2018 August 20, we use \textit{Global Oscillation Network Group (GONG)}/H$_{\alpha}$ observation and high-resolution EUV imaging filters. Careful inspection offers different possibilities of CME2's source region, which are associated with the continuous drainage of filament, coronal plasma channel, and overlying flux-rope. The filament is shown in panels a, b, and c of Figure (\ref{fig:multi-wavelength_CME2}) lying between the two trans-equatorial coronal holes (panels d, e, and f of Figure \ref{fig:multi-wavelength_CME2}). An elongated hot coronal plasma channel/filament channel associated with this filament appears in the AIA coronal channels at 2-4 MK (193 {\AA} and 211 {\AA}). We could see the drainage of the filament and associated elongated plasma channel erupt around $\approx$ 08:00 UT. The filament eruption occurs slowly, while the coronal plasma channel erupts at a higher speed \citep{Chen2019,Gopalswamy2022, Mishra2019, Palmerio2022}. However, there is no signature on the solar disk during both eruptions, indicating the stealthy nature of this eruption \citep{Nitta2021,Mishra2019, Palmerio2022}. But, several hours later, the formation of post-eruption arcades and coronal dimming indicates the earlier eruption, which possibly appeared as an overlying flux-rope above the coronal plasma channel. After careful inspection, this overlying diffuse flux-rope is visible in STEREO-A/EUVI-171 {\AA} waveband (panels g, h, and i of Figure  \ref{fig:multi-wavelength_CME2};). We note a small portion of the filament lying over the coronal hole erupted in a jet-like fashion around $\approx$19:00 UT \citep{Mishra2019,Palmerio2022} together with the overlying flux-rope and collectively led to CME2. It is clear that CME2 resembles a stealth CME and shows no obvious signatures of having a hot flux-rope contrary to CME1. Further, the kinematics of this faint CME1 does not show its fast expansion in the higher corona. The absence of faster cooling and hot flux-rope for CME2 implies that it is unlikely to release heat in the corona. This is consistent with findings from the FRIS model that the polytropic index is not higher than the adiabatic index for CME2 during its tracking between 5-15 $R_\odot$.

\section{Discussion}\label{sec:discussion}

Our study has unveiled the disparity in the evolution of the thermal properties and the internal dynamics of two CMEs (CME1: 2011 September 24 \& CME2: 2018 August 20) significantly differing in their kinematics at lower coronal heights.  CME1 is a high-speed (1,885 km s$^{-1}$ ) CME with two-phase kinematics, a rapid decrease in expansion rate at the beginning (within 8.4 $R_\odot$ ) followed by a nearly constant expansion in the later phase (up to 20 $R_\odot$) of its journey. In contrast, CME2 has a relatively slow speed (420 km s$^{-1}$), showing only a gradual increase in the expansion rate throughout (5 to 15 $R_\odot$) our observations. Investigating the thermodynamics of CMEs, we find differential heating of the selected CMEs: CME1 in its early-rapid decrease in acceleration phase shows heat-release, but CME2 in its gradual-acceleration phase shows heat-absorption. However, both the CMEs during their later propagation phase indicate heat-absorption, making them closer to the isothermal state than the adiabatic state. By seeing the trend in the polytropic index value for CME2, it can be inferred that CME2 might also release heat at the beginning of its journey. Interestingly, the fast-speed CME1 reaches the isothermal state ($\Gamma=1$) rapidly and at a lower height than the slow-speed CME2. The thermodynamic evolution of both the selected CMEs validates one compelling finding: CMEs are not going through ideal adiabatic expansion but rather experience a heat-release and/or heat-absorption state during their heliospheric propagation.

We find that centrifugal and thermal pressure forces primarily contribute to the expansion of the CMEs near and far from the Sun, respectively, but the Lorentz force prevents the expansion throughout. In the early-rapid decrease in the acceleration phase of the CME1, the rate of decrease is highest for thermal pressure force, intermediate for Lorentz force and the lowest for centrifugal force. However, in the gradual acceleration phase, the rate of decrease is the highest for centrifugal force, intermediate for Lorentz force and the lowest for thermal pressure force. Thus, considering a higher magnitude at initial heights, the contribution to radial expansion is more by the centrifugal force. The slowest decreasing trend in thermal pressure force suggests its dominance towards the radial expansion at higher heights. One interesting trend is that the thermal pressure force is already overtaking the centrifugal force and is near to overtaking the Lorentz force for the fast speed CME1 at the height of 18 $R_\odot$. On the other hand, this possibility is not yet achieved by the CME2 in our observed heights. Thus, by seeing the trend, it can be inferred that the dominance of thermal pressure force over the other two for the slow speed CME2 may be attained at a higher height than the fast speed CME1.

In our study, the inferences from multi-wavelength observations are consistent with our model-derived findings at initially observed heights. The initial heat-release in CME1 is expected as it is heated to 8-15 MK during its eruption itself. However, we found no obvious reason for heat-absorption in CME2 at initial heights. It is possible that the presence of a coronal hole in the vicinity of the source region and the continuous fractional eruption of the filament can put some heat into the CME2 flux-rope. Previous studies for this event also reported that the magnetic reconnection at the source region powered the continuous positive acceleration\citep{Gopalswamy2022}. Thus, magnetic reconnection could be a physical mechanism for the observed heating in our model-derived result for CME2. The selected CMEs could not be tracked during their expected early-rise and impulsive acceleration much closer (within 2-3 $R_\odot$) to the Sun. Therefore, our findings further need to be examined at heights that are most often unobserved due to coronagraphic occulter and/or projection effects on the Earth-directed CMEs. We aim to address this limitation using coronagraphic observations of limb CMEs from Mauna Loa Solar Observatory (MLSO) and upcoming \textit{Aditya-L1}. It is noteworthy that we did not attempt to identify the physical processes leading to differential heating of the selected CMEs.

The physical mechanisms responsible for heating the CMEs are poorly understood, and there could be many probable reasons for this. It might be possible that the heat is transferred from the lower corona to the CME flux-rope as its foot-point is believed to be connected to the Sun \citep{Larson1997}. Another possibility is a continuous injection of thermal radiation energy from the solar wind into CME in the outer corona. However, transporting charge particles in directions perpendicular to the average magnetic field (cross-field diffusion) will be limited \citep{Zhang2003}. The heating of CMEs will also be observed if the magnetic energy of the CME dissipates at varying rates at different heights, but one needs to identify processes for such dissipation and its effect on global CME kinematics. It is found that untwisting of the flux-rope can provide the required energy for the propagation and expansion of the CME \citep{Vourlidas2000}, while heat could be generated at the expense of a fraction of the internal magnetic energy \citep{Kumar1996}. Such a conversion of the magnetic energy to thermal energy is by the Joule heating ($j^2/\sigma$) process, where $j$ is the current density and $ \sigma$ is the electrical conductivity. However, a very high value of $\sigma$ in the interplanetary medium suggests that joule heating cannot be significant. Additionally, the possibility of internal magnetic reconnection in CMEs \citep{Farrugia1993} or the reconnection between CMEs and interplanetary magnetic field (IMF) \citep{Lugaz2013} might play an important role in converting magnetic energy into thermal energy for supplying heat to the plasma embedded in the CME flux-rope. Our study has the limitation of unveiling the responsible physical processes for estimated heating for selected CMEs. Therefore, future studies should focus on quantifying the heat-generation efficiency of different mechanisms in the CMEs.

Our findings are primarily based on distance-dependent variations in the polytropic index, heating rate, temperature and internal forces of CMEs, estimated by implementing the revised FRIS model on the CME's 3D kinematics. The derived FRIS model with needed corrections is described in the appendix (\ref{appendix:derivation}), and the revamped thermodynamic parameters are shown in Table (\ref{tab:parameters}). The 3D kinematics, as shown in Figure ({\ref{fig:kinematics}}), is estimated by implementing the GCS model to the coronagraphic observations from multiple viewpoints. Using models and multi-wavelength flux-rope CME observations, our study sheds light on the heating anomalies in CMEs differing significantly in their kinematic behavior.

Our study establishes a crucial link between the diverse kinematics of coronal mass ejections (CMEs) and their thermodynamic evolution at heights where thermodynamic measurements are scarce. By employing global kinematics, we implicitly account for the effects of CMEs interacting with the ambient medium without imposing constraints on force-free or non-force-free flux-ropes in our FRIS model output. However, we acknowledge potential uncertainties in the estimated thermal properties due to observational limitations and model considerations. Moreover, fitting coefficients are averaged over the entire propagation period, which may introduce additional uncertainties. While multi-wavelength observations can aid in comprehending the thermal evolution of CMEs, validating the model output is challenging without direct in-situ measurements of thermodynamic parameters. To address this, it is essential to analyze CMEs observed in-situ by the \textit{Parker Solar Probe} (PSP) and \textit{Solar Orbiter} (SolO) at unprecedentedly close distances to the Sun, along with multiple in-situ spacecraft at varying radial distances in the interplanetary medium. This will allow for continuous estimation of thermodynamic parameters and facilitate comparison with direct in-situ measurements. We demonstrate that understanding the thermodynamic behavior of CMEs remains enigmatic, requiring the utilization of multiple independent models, multi-wavelength imaging, and in-situ observations to elucidate the mechanisms responsible for the observed differential heating in CMEs. By pursuing these avenues, we can unravel the mysteries surrounding CME thermodynamics and advance our comprehension of these dynamic solar events.

\begin{acknowledgments}
 We acknowledge the instruments team members for providing the STEREO (EUVI, COR1, and COR2), SOHO (LASCO), and SDO (AIA and HMI) data. 
\end{acknowledgments}

\facilities{SOHO (LASCO C2 and C3), STEREO (EUVI, COR1 and COR2), SDO (AIA and HMI), and GONG/H$_\alpha$}


\appendix

\section{Derivation of Flux-Rope Internal State (FRIS) Model}\label{appendix:derivation}
Unlike on the global scale, where the CMEs may be a loop-like structure with two ends rooted on the surface of the Sun, on the local scale, we may assume them to be an axisymmetric cylindrical flux-rope. In the cylindrical coordinates$(r, \phi, z)$ with the origin on the axis of the flux-rope, we will have $\frac{\partial }{\partial \phi }=\frac{\partial }{\partial z } =0$. The cross-section of the flux-rope is assumed to be a circle with a radius $R(t)$, which is a time-dependent parameter. The expansion speed and the expansion acceleration, respectively, are given by $v_e( t )= \frac{dR(t)}  {dt}$ and $a_e( t )= \frac{dv_e(t)}  {dt}$, at the boundary of the flux-rope.

Considering a self-similar expansion of the flux-rope, mathematically, we can define a normalised radial distance $x$ from the axis of the flux-rope that will be independent of time and is given by $x = \frac{r}  {R} $, $x=1$ is the boundary of the flux-rope. Now the radial and azimuthal components of the propagation velocity can be written in terms of $x$ as

\begin{equation}
 v_r( t,x ) = \frac{dr}  {dt} =x \frac{dR(t)}  {dt}=xv_e( t )\label{eqn:Vr}
\end{equation}
\begin{equation}
  v_\phi ( t,x )=r \frac{d\phi(t)}  {dt}=f_{v\phi }( x ) v_p( t )\label{eqn:VrVphi}    
\end{equation}
Where $v_p$ is the poloidal speed at the boundary of the flux-rope, and $f_{v\phi }( x )$ is an unknown distribution function for the azimuthal component of the propagation velocity.

\subsection{Conservations of Total Mass and Angular momentum}
Considering the magnetic field lines are frozen in with the plasma flows and the self-similar expansion, the flux-rope will have a fixed distribution $f_p(x)$ of density in the radial direction. So the density in the flux-rope can be written as $\rho( t,x ) =f_\rho ( x ) \rho_{0}( t )$, where $\rho_{0}( t ) = \bar {\rho} = \frac{M}  {\pi l R^2}  $ is the average mass density of the flux-rope, and l is the axial length of the flux-rope.

The total mass of a CME can be written as 
\begin{equation}
\begin{array}{rl}
   &M = \int_V \rho \,dV=\int_V \rho \,rdrd\phi dz {\nonumber}\\
   &\Rightarrow M = 2\pi R^2 l \rho_{0}(t) \int_{0}^{1}   f_\rho( x ) \,xdx
   \end{array}
\end{equation}

The total angular momentum of the flux-rope is given by
\begin{equation*}
   \begin{array}{rl}
  & L_A = \int_V \rho rv_\phi \,rdr d\phi dz\\ 
  & \Rightarrow L_A = 2MR v_p(t) \int_{0}^{1} {f_\rho(x) f_{v\phi}(x) x^2} \,dx
   \end{array} 
\end{equation*}
 \begin{equation}
  \Rightarrow v_p ( t )= \frac{L_A}  {2MR \int_{0}^{1} {f_\rho(x) f_{v\phi}(x) x^2} \,dx } \label{eqn:vp}
\end{equation}

Putting the equation (\ref{eqn:vp}) in equation (\ref{eqn:VrVphi}), we will have
\begin{align}
\label{eqn:vphi}
v_\phi (t,x )= f_{v\phi} ( x ) v_p ( t )\implies v_\phi ( t,x) = \frac{L_A k_1 } { MR } f_{v\phi} ( x )
\end{align}  
where, $k_1 = \frac{1}  {2\int_{0}^{1} f_\rho(x) f_{v\phi}(x) x^2 \,dx } $ is an unknown integral constants.

\subsection{Equations of Thermodynamics}

For ideal gas and reversible processes, the first and second laws of thermodynamics are given by, respectively,
\begin{align}
\label{eqn:thermo}
    du = dQ - pd\left(\frac{1}{\rho}\right)\text{ and}\hspace{0.1cm}ds = \frac{dQ}{T}
\end{align}
where, $u = \frac{p}{(\gamma-1)\rho}$, $p = nk(T_p+T_e) = 2nkT$, $ n = \frac{\rho }{m_p}$ are the internal energy per unit mass, thermal pressure and number density of proton respectively. $\rho$ is the mass density of protons, $m_p$ is the proton mass, $\gamma$ is the adiabatic index which is 5/3 for monoatomic ideal gases. $T_p$ and $T_e$ are the temperatures of the proton and electron, respectively. 

Solving equation (\ref{eqn:thermo}), we will get
\begin{align*}
    & ln \left( \frac{p}{{\rho}^{\gamma}} \right) =\frac {  (\gamma -1)T \rho s}{ p }= \frac{  (\gamma -1)m_p s}{ 2k } \\
    &\Rightarrow  p = e^{\left[\frac {(\gamma -1)m_p s}{2k} \right]}\rho ^{\gamma}
\end{align*}
putting \hspace{0.2cm}  $\frac{(\gamma -1)m_p }{2k} = \sigma =\text{constant}$ \hspace{0.1cm} in the above equation, we will get \hspace{0.5cm} 
\begin{equation}\label{eqn:pres}
    p(t,r)=e^{ \sigma s } \rho^{\gamma}
\end{equation}

\subsection{Internal Forces}

The Magnetohydrodynamic equation of motion for a magnetised fluid element is given by
\begin{equation}
\label{eqn:MHD}
    \rho \left( \frac{\partial \vec{V}}{\partial t} \right) + \rho (\vec{V} \cdot \nabla) \vec{V} = - \nabla p + \vec{j} \times \vec{B}
\end{equation}
where, p is the plasma thermal pressure, $\vec{B}=(0,B_\phi,B_z)$ is the magnetic field and $\vec{j}=  \frac{\nabla \times \vec{B}}{\mu_o}$ is the current density. The equation (\ref{eqn:MHD}) represents the plasma's equation motion in the inertial frame of the flux-rope. Thus it describes the expansion behaviour of the flux-rope.

The radial component of the MHD equation (\ref{eqn:MHD}) is given by  
\begin{equation*}
   \rho \frac { \partial v_r}{ \partial t}+\rho \left({ v_r \frac{\partial v_r}{\partial r}}-\frac{{v_\phi}^2}{r}\right)=- \frac{ \partial p}{\partial r} + (\vec {j} \times   \vec {B} )_r
\end{equation*}

By using equations (\ref{eqn:Vr}) and (\ref{eqn:vphi}) in the above equation, we get
\begin{equation}
\label{eqn:fem_r}
    ( \vec{j} \times \vec {B} )_r =\rho \left({a_e x - \frac{{k_1}^2 {L_A}^2 {f_\phi}^2}{M^2 x R^3} }\right) +\frac{ \partial p}{\partial r }
\end{equation}
By integrating equation (\ref{eqn:fem_r}) over the cross-section of the flux-rope, we can calculate the average Lorentz force and is given by
\begin{align}
   &{{\bar f}_{em}} =\frac{1}{ \pi R^2} \int_{0}^{R} \int_{0}^{2\pi} ( \vec{j} \times \vec {B} )_r \,rdr \,d\phi{\nonumber}\\
    &={2} \int_{0}^{1} {\rho_0 f_\rho(x)(a_e x - \frac{k_1 ^2 L_A ^2 f_\phi ^2}{M^2 x R^3})} \,xdx + \frac{ 2}{R } \int_{0}^{1} {x \frac{\partial p}{\partial x}}\,dx
\end{align}
Using equation (\ref{eqn:pres}) and $\rho_0 = \frac{M}{ \pi l R^2} $ in the above equation and by solving, we will get
\begin{align}
\label{eqn:fem}
   {{\bar f}_{em}} = \frac{ k_2 M a_e}{lR^2 }-\frac{ k_1^2 k_3 L_A^2}{MlR^5 }-{\frac{M^\gamma e^{\sigma s}}{l^\gamma R^{2\gamma +1}} } k_4
\end{align}
Where, $k_2 = \frac{ 2}{\pi } \int_{0}^{1} {f_\rho( x ) x^2} \,dx>0, k_3 =\frac{ 2}{\pi } \int_{0}^{1} {f_\rho( x )f_{v\phi}^2( x )} \,dx\ge 0 $ and $k_4 =\frac{2}{\pi^\gamma}\left[  \int_{0}^{1} {f_\rho^\gamma(x)}\,dx - f_\rho^\gamma(1)\right]$ are unknown integral constants. The first, second and third terms on the right-hand side of the equation (\ref{eqn:fem}) are the average total force due to the expansion, the average force due to poloidal motion, and the average thermal pressure force, respectively.
The magnetic field for an axisymmetric cylindrical flux-rope can be written as
\begin{equation*}
\vec{B} = B_\phi \hat{\Phi} + B_z  \hat {z} = \nabla \times \vec {A} \hspace{0.3cm}  \text{with} 
\end{equation*}
\begin{equation*}
B_\phi = - \frac{ \partial A_z}{\partial r }\text{and} \hspace{0.1cm} B_z =\frac{ 1}{r }\frac{\partial(rA_\phi) }{ \partial r}
\end{equation*}
Under the assumption of self-similar expansion, the magnetic fluxes are conserved in $\phi$ and $z$ direction and can be written as
\begin{equation}
\label{eqn:magFlux}
\left.
  \begin{array}{rl}
&F_\phi = -l \int_{0}^{R} {\frac {\partial A_z}{\partial r}}\, dr =  l\left([A_z(0)-A_z(R)\right]\\
&F_z = 2\pi \int_{0}^{R} { \frac{\partial( r A_\phi)}{\partial r} }\,dr =  2\pi RA_\phi(R)
\end{array}
\right\}  
\end{equation}
The magnetic vector potentials $A_\phi$ and $A_z$ also conserve their own distributions and can be expressed as
\begin{equation}
\label{eqn:magVec}
\left.
  \begin{array}{rl}
&A_\phi = f_{A\phi}( x )A_{ \phi o }( t )\\
&A_z = f_{Az}( x )A_{ z o }( t )
\end{array}
\right\}  
\end{equation}

Solving equations (\ref{eqn:magFlux}) and (\ref{eqn:magVec}), we will get
\begin{equation}
\label{eqn:magVec2}
 A_\phi(t,x) = \frac{ f_{A\phi}(x)}{R }\hspace{0.1cm}\text{and}\hspace{0.1cm}A_z (t,x) = \frac{f_{Az}( x )}{l}  
\end{equation}
Using equations (\ref{eqn:magVec}) and (\ref{eqn:magVec2}), we can find out the average Lorentz force over the cross-section of the flux-rope as 
\begin{align}
\label{eqn:fem2}
 {{\bar f}_{em}} &=\frac{1}{ \pi R^2} \int_{0}^{R} \int_{0}^{2\pi} ( \vec{j} \times \vec {B} )_r \,rdr \,d\phi=\frac{2}{ R^2} \int_{0}^{R} {( \vec{j} \times \vec {B} )_r} \,rdr {\nonumber}\\
&=\frac{2}{ R^2}\int_{0}^{R}  {\biggl[ \frac{ 1} {\mu _0 }(-\frac{\partial B_z}{\partial r} \hat{\phi} +\frac{\partial rB_\phi}{\partial r} \hat {z})\times (B_\phi \hat {\phi} + B_z \hat {z}) \biggr]_r} \,rdr{\nonumber}\\
 \Rightarrow {{\bar f}_{em}}&={-\frac{k_5 } {\mu_0 R^5 } }-{ \frac{k_6}{\mu_0 l^2 R^3} }
\end{align}
Where, $k_5 =2 {\int_{0}^{1} {[\frac{\partial( x f_{A \phi})}{ \partial x}][\partial \frac{(\frac{\partial(x f_{A \phi})}{x \partial x})}{\partial x}]}\,dx}$ and $k_6 ={2 \int_{0}^{1}{[\frac{\partial f_{Az}}{ \partial x}][\frac{\partial (x\frac{\partial f_{Az}}{ \partial x})}{\partial x}]}\,dx}$ are unknown integral constants. 

Using the current observations, it is difficult to accurately measure the flux-rope's axial length (l). Under the assumption that the axial length (l) is proportional to the distance (L) between the axis of the flux-rope and the solar surface, where L can be measured using imaging observations of CMEs, we can write
\begin{equation}
    \label{eqn:L}
    \text{ $l=k_7 L$\hspace{0.3cm}, where $k_7$ is a positive constant.}
\end{equation}
Using equation (\ref{eqn:L}) and equating equations (\ref{eqn:fem}) and (\ref{eqn:fem2}), we will get
\begin{equation}
\label{eqn:motion1}
    a_e - c_1 R^{-3} = c_2 LR^{-3} + c_3 L^{-1} R^{-1} + \lambda ( t ) L^{ 1-\gamma } R^{ 1-2\gamma }
\end{equation}
This is the equation for the radial expansion of the flux-rope, where $c_1 = \frac{ k_1^2 k_3 L_A^2}{k_2 M^2 }\geq 0 $, $c_2 = \frac{ -k_5 k_7}{\mu_0 k_2 M }$, and $ c_3=\frac{  -k_6}{\mu_0 k_2 k_7 M }\leq 0$ are constants. $\lambda(t)= c_0 e^{ \sigma s(t) }$ is a time depedent parameter with $c_0 = \frac{ k_4 M^{\gamma -1}}{k_2 k_7^{\gamma -1} }$ is a constant.
The first and second term in the left-hand side of the equation (\ref{eqn:motion1}) gives the acceleration due to radial and poloidal motion, respectively. The first two terms on the right-hand side of the equation (\ref{eqn:motion1}) represent the Lorentz force, and the third term represents the thermal pressure force.  

Thus, the average Lorentz force, the average thermal pressure force, and the average centrifugal force can be written in terms of the measurable parameters (L, R), the unknown constants $c_1-c_3$, and $\lambda(t)$ as
\begin{align}
    &{{\bar f}_{em}} ={-\frac{k_5 } {\mu_0 R^5 } }-{ \frac{k_6}{\mu_0 l^2 R^3} }= \frac{k_2 M}{ k_7 }{[ {c_2 R^{-5}}+{c_3 L^{-2} R^{-3}} ]}\label{eqn:fem_final}\\
    &{{\bar f}_{th} }= \frac{k_4 M^\gamma e^{ \sigma s  }} { l^\gamma R^{2\gamma +1}} = {\frac{ k_2 M}{k_7 }}{ [\lambda(t) L^{-\gamma} R^{-2\gamma -1}] }\label{eqn:fth_final}\\
    &{{\bar f}_p} = \frac{k_1^2 k_3 L_A^2 } {MlR^5} = {\frac{k_2 M}{ k_7}}{ [c_1 R^{-5} L^{-1}] }\label{eqn:fp_final}
\end{align}

\subsection{Density, Temperature and Thermal pressure}

The average proton mass density, the average proton number density, the average thermal pressure, and the average temperature, respectively, can be expressed as
\begin{align}
    &\bar {\rho} = \frac{M  }  { \pi l R^2 } =\frac{M  }  { \pi k_7  } ( L R^2 )^{-1} \\
    &{{\bar n}_p} =\frac{\bar \rho} {m_p}  = \frac{M  }  { \pi k_7 m_p  } ( L R^2 )^{-1}\label{eqn:np}\\
    &{\bar p}( t ) = \frac{ 2}{ R^2} \int_{0}^{R} p \,rdr = \frac{ k_2 k_8 M  } {k_4 k_7  }{ \lambda (LR^2)^{-\gamma} }\hspace{0.3cm}\\
    &\text{where, } k_8 = {\frac{2} {\pi ^\gamma }\int_{0}^{1}  {f_\rho ^\gamma(x)}   \,xdx}{\nonumber}\\
    &\bar {T }=\frac{\bar {p} } { 2 \bar {n_p} k}=\frac{\pi \sigma}   {(\gamma -1)} \frac{ k_2 k_8}{k_4 }{\lambda (LR^2)^{1-\gamma}  }\label{eqn:T}
\end{align}

\subsection{Polytropic Index }
During the propagation, the CME may go through a polytropic process with multiple expansion and compression, which include heat transfer. The polytropic processes provide a novel way to study plasma thermodynamics through the relationship between macroscopic parameters. A polytropic process obeys the relation between a ﬂuid’s thermal pressure (p) and density ($\rho$) with an index $\Gamma$ described as $p = b(t){\rho}^{\Gamma}$.
So putting this value in the equation (\ref{eqn:pres}), we will get
\begin{equation}
\label{eqn:sigma}
     e^{ \sigma s }=b{\rho} ^{\Gamma-\gamma} 
\end{equation} 

Using equation (\ref{eqn:sigma}), the expression for $\lambda(t)$ can be written as
\begin{align}
  \label{eqn:lamb1}
    &\lambda( t ) = c_o e^{ \sigma s } =c_o b( t ) \rho^{ \Gamma -\gamma }{\nonumber}\\
    &\Rightarrow  \Gamma = \gamma + \frac{ln[\lambda(t)] -ln[c_o b(t)]}  {ln[\rho(t)]} 
\end{align}
Assuming the thermodynamic process to be quasi-static in between two measurement points i:e between $t$ and $t + \Delta t$, where $\Delta t$ is a relatively small interval compared to the whole time scale of measurements. So the equation (\ref{eqn:lamb1}) can be written at time  $t + \Delta t$ as 
\begin{equation}
\label{eqn:lamb2}
    \Gamma \approx \gamma + \frac{ln[\lambda(t+\Delta t)] -ln[c_o b(t\Delta t)]}  {ln(\rho(t))} 
\end{equation}
Where the values of $\Gamma$ and $b$ are nearly constant during the $\Delta t$ time interval.
Solving equations (\ref{eqn:lamb1}) and (\ref{eqn:lamb2}), we will get the final expression for the polytropic index as,
\begin{equation}
\label{eqn:lamb}
    {\Gamma} \approx \gamma +  { \frac{ln\left[{\frac{\lambda (t)}  {\lambda (t+\Delta t)}}\right]} {ln\left[\frac{L(t+\Delta t)}{L(t)}\left(\frac{R(t+\Delta t)}  {R(t)}\right)^2\right]} }
\end{equation}

\subsection{Rate of change of Entropy and Heating Rate}
The rate of change of entropy per unit mass of the CME can be derived using the expression of $\lambda(t)$ as follows,
\begin{align}
    \label{eqn:entropy}
    &\lambda( t ) = c_o e^{ \sigma s } {\nonumber}\\
    &\Rightarrow \frac{d\lambda }{dt } = {\sigma \lambda}\frac{ds }{dt }{\nonumber}\\
    &\Rightarrow \frac{ds }{dt } = \frac{1}{{\sigma \lambda}}\frac{d\lambda }{dt } 
\end{align}

The average heating rate per unit mass of the CME can be written as
\begin{align}
    \label{eqn:heating}
   &\bar {\kappa} ( t ) = \frac{dQ(t) }{dt }  = \bar{T} \frac{ ds} {dt } {\nonumber}\\
   &\Rightarrow \bar {\kappa} ( t )) =\frac{\pi } {(\gamma -1)} \frac{ k_2 k_8}{k_4 }{ (LR^2)^{1-\gamma}  }\frac{d\lambda }{dt }
\end{align}

\subsection{Equation of radial expansion of the CME }
The equation of motion (\ref{eqn:motion1}) can be written as
\begin{equation}
\label{eqn:motion2}
    \lambda ( t )= L^{ \gamma -1 } R^{ 2\gamma -1 }( a_e - c_1 R^{-3} - c_2 LR^{-3} - c_3 L^{-1} R^{-1} )
\end{equation}
Only using the measurable parameters L, R, and their derivatives, the equation (\ref{eqn:motion2}) is not sufficient to calculate the unknown parameters $c_1-c_3$ and $\lambda (t)$. So we need an additional constraint on $\lambda (t)$. Now assume that irrespective of any heating mechanism in the CMEs, the heating rate  per unit mass may be equivalently expressed as the result of heat flow. So
\begin{align*}
 &{\bar {\kappa}}  \propto {\frac{Area \cdot \Delta T}{length} }\Rightarrow {\bar {\kappa}} = k_{11}{\frac{T_a - {\bar{T}}}{{\bar{\rho}}L^2 }} \\
 &\Rightarrow \frac{ \pi k_2 k_8}{(\gamma -1)k_4 }(LR^2)^{1-\gamma}\frac{d\lambda }{dt}= k_{11} \left[ \frac{{T_a}-{\frac{\pi \sigma k_2 k_8}{(\gamma -1)k_4}\lambda(LR^2)^{1-\gamma}}} {\frac{M}{\pi k_7} (LR^2)^{-1} L^2} \right]
\end{align*}
\begin{equation}
\label{eqn:motion3}
 \Rightarrow(LR^2 ) ^{ \gamma -1 } = c_4 LR^{-2} \frac{d\lambda }{dt } + c_5 \lambda 
\end{equation}
 Where $k_{11}$ is an unknown constant, $T_a$ could be treated as the average temperature of the ambient solar wind, $c_4 = \frac{ k_2 k_8 M}{(\gamma -1) k_4 k_7 k_{11} T_a}$ and $c_5 = \frac{\pi \sigma k_2 k_8 }{ (\gamma -1) k_4 T_a}$ are unknown constant integrals.

On integrating equation (\ref{eqn:motion2}), we will get 
\begin{align}
\label{eqn:dlambda}
 \frac{ d\lambda}{dt }=L^{ \gamma -1 }R^{2 \gamma -1 }& \biggl[\frac{da_e}{dt}+{(2\gamma -1){a_e v_e R^{-1}}+(\gamma -1) a_e v_c L^{-1}}\biggr.{\nonumber}\\
 &\left.+ c_1 ({(4-2\gamma)v_e R^{-4}}+{(1-\gamma)v_c L^{-1} R^{-3}})\right.{\nonumber}\\
 &\left.+ c_2 ({(4-2\gamma )v_e L R^{-4} - {\gamma v_c R^{-3}}})\right.{\nonumber}\\
 &\biggl.+ c_3 ({( 2-2\gamma ) v_e L^{-1} R^{-2}} +(2-\gamma) v_c L^{-2} R^{-1})\biggr] 
\end{align}

Now putting equation (\ref{eqn:motion2}) and (\ref{eqn:dlambda}) in equation (\ref{eqn:motion3}) and solving, we will get
\begin{align}
(LR^2)^{\gamma -1} = &L^{\gamma -1} R^{2\gamma -1}\biggl[ ({c_5 a_e}-{c_3 c_5 L^{-1} R^{-1}}+({{c_4 L \frac {da_e}{dt}}+{(\gamma -1)c_4 a_e v_c}}) R^{-2} )  \biggr.{\nonumber}\\
&\left.+  ({(2-\gamma)c_3 c_4 v_c L^{-1}}+{(2\gamma -1)c_4 a_e v_e L}-{c_2 c_5 L}-{c_1 c_5}  )R^{-3} + ((2-2\gamma)c_3 c_4 v_e  )R^{-4} \right.{\nonumber}\\
&\biggl.+ ({(1-\gamma)c_1 c_4 v_c}-{\gamma c_2 c_4 v_c L} )R^{-5} + ({(4-2\gamma)c_1 c_4 v_e L}+{(4-2\gamma)c_2 c_4 v_e L^2})R^{-6}\biggr]
\end{align}

Rearranging the above equation, we can write it further as the following. 

\begin{align}
\frac {R}{L} =  &{c_5 \biggl[\frac {a_e R^2}{L}\biggr]}-{c_3 c_5 \biggl[\frac {R}{L^2}\biggr]}-{c_2 c_5\biggl[\frac {1}{R}\biggr]}-{c_1 c_5\biggl[\frac {1}{LR}\biggr]} {\nonumber}\\
&+c_4 \biggl[\frac {da_e}{dt}+{\frac {(\gamma -1)a_e v_c}{L}}+\frac{(2\gamma -1)a_e v_e}{R}\biggr] + c_3 c_4\biggl[ \frac{(2-\gamma) v_c }{L^2 R}+{\frac{(2-2\gamma) v_e }{LR^2}}\biggr] {\nonumber}\\ 
&+c_2 c_4\biggl[{\frac{(4-2\gamma) v_e L}{R^4}}-{\frac{\gamma v_c}{R^3}}\biggr] + c_1 c_4\biggl[{\frac{(4-2\gamma) v_e}{R^4}}+{\frac{(1-\gamma)v_c}{LR^3}}\biggr]
\label{fitting}
\end{align}

The above equation (\ref{fitting}) is the final equation of motion for the radial expansion of the CME. The unknown constants $c_1 - c_5$ and $\lambda(t)$ can be obtained by fitting the equation (\ref{fitting}) to the measurements of L, R and their time derivatives.

\section{The GCS model-fitted parameters for CME1 and CME2 \label{appendix:gcs_parameters}}
 The GCS model has six parameters (longitude, latitude, leading edge height, aspect ratio, tilt angle and half angle) to reproduce the 3D structure of a CME. We have fitted the contemporaneous images from three vantage points in each time step. The GCS parameters for all the time steps in our observation for both the selected CMEs, CME1 and CME2, are listed in Table (\ref{tab:CME1_gcs_all}) and (\ref{tab:CME2_gcs_all}), respectively. We found the longitude, latitude, aspect ratio, tilt angle and half angle for CME1 to be -41$^{\circ}$, 13$^{\circ}$, 0.39, -62$^{\circ}$, and 26$^{\circ}$, respectively. The obtained longitude, latitude, aspect ratio, tilt angle and half angle for CME2 were 10$^{\circ}$, 5$^{\circ}$, 0.27, 10$^{\circ}$, and 16$^{\circ}$, respectively. The obtained model-fitted parameters, except the leading-edge height, remain the same during each time step of the observed evolution phase for both CMEs. Thus, no deflection and rotation of the selected CMEs are found during their propagation. Further 
 both the CMEs follow a self-similar evolution as there is no change in half angle and aspect ratio for each time step during our observations.

\begin{table}[ht]
\parbox{.45\linewidth}{
\centering
\caption{\label{tab:CME1_gcs_all}The GCS model-fitted leading-edge height for CME1 (2011 September 24) at each time step during our observation.}
\begin{tabular}{ccc}
\hline
\textbf{Date} &\textbf{Time} & \textbf{Height} \\
 & \textbf{(UT)} &   \textbf{($R_\odot$)} \\
\hline
\hline
2011 September 24 & 12:45 & 2.5 \\
\hline
2011 September 24 & 12:50 & 3.2\\
\hline
2011 September 24 & 12:55 & 4.0\\
\hline
2011 September 24 & 13:24 & 8.3\\
\hline
2011 September 24 & 13:30 & 9.2\\
\hline
2011 September 24 & 13:39 & 10.6\\
\hline
2011 September 24 & 13:54 & 12.9\\
\hline
2011 September 24 & 14:06 & 14.8\\
\hline
2011 September 24 & 14:18 & 16.7\\
\hline
2011 September 24 & 14:24 & 17.7\\
\hline
2011 September 24 & 14:30 & 18.6\\
\hline
2011 September 24 & 14:39 & 20.1\\
\hline
\end{tabular}
}
\hfill
\parbox{.45\linewidth}{
\centering
\caption{\label{tab:CME2_gcs_all}The GCS model-fitted leading-edge height for CME2 (2018 August 20) at each time step during our observation.}
\begin{tabular}{ccc}
\hline
\textbf{Date} &\textbf{Time} & \textbf{Height} \\
 &\textbf{(UT)} &   \textbf{($R_\odot$)} \\
\hline
\hline
     2018 August 20 &    21:24 &     5.7 \\ 
      \hline
     2018 August 20 &   22:24  &    6.4 \\
      \hline
     2018 August 20 &   23:24 &    7.1 \\
      \hline
     2018 August 20 &   23:39  &    7.2 \\
      \hline
     2018 August 20 &   23:54  &    7.4 \\
      \hline
     2018 August 21 &   0:24  &    7.7 \\
      \hline
     2018 August 21 &   0:39  &    7.9 \\
      \hline
     2018 August 21 &   0:54  &    8.1 \\
      \hline
     2018 August 21 &   1:24  &    8.5 \\
      \hline
     2018 August 21 &   1:39  &    8.7 \\
      \hline
     2018 August 21 &   1:54  &    8.8 \\
      \hline
     2018 August 21 &   2:24  &    9.3 \\
      \hline
     2018 August 21 &   2:39  &    9.5 \\
      \hline
     2018 August 21 &  2:54  &    9.8 \\
      \hline
     2018 August 21 &   3:24  &    10.4 \\
      \hline
     2018 August 21 &   3:39  &    10.7 \\
      \hline
     2018 August 21 &   3:54  &    11.0 \\
      \hline
     2018 August 21 &   4:24  &    11.7 \\
      \hline
     2018 August 21 &  4:39  &    12.1 \\
      \hline
     2018 August 21 &  4:54   &   12.4 \\
      \hline
     2018 August 21 &   5:24   &   13.2 \\
      \hline
     2018 August 21 &  5:39  &    13.6 \\
      \hline
     2018 August 21 &  5:54  &    14.1 \\
      \hline
     2018 August 21 &   6:24  &    15.2 \\
\hline
\end{tabular}
}
\end{table}

\vspace{-35pt}


\end{document}